\documentclass{aa}

\usepackage[varg]{txfonts}
\usepackage{amssymb}
\usepackage{epsfig}
\usepackage{graphics}
\usepackage{amsmath}
\usepackage{color}
\usepackage{natbib}
\usepackage{mathrsfs}
\usepackage{afterpage}
\usepackage{amsfonts}
\usepackage[normalem]{ulem}

\bibpunct{(}{)}{;}{a}{}{,} 

\DeclareMathAlphabet{\mathbbmsl}{U}{bbm}{m}{sf}
\DeclareMathAlphabet{\mathpzc}{OT1}{pzc}{m}{it}

\newcommand{\be}{\begin{equation}}
\newcommand{\ee}{\end{equation}}
\newcommand{\bea}{\begin{align}\begin{split}}
\newcommand{\eea}{\end{split}\end{align}}

\newcommand{\Msun}{\ensuremath{{M}_{\sun}}} 
\newcommand{\grg}{\ensuremath{g_{\mathrm{rad}}/g}}
\newcommand{\lognat}{\ln}

\newcommand{\clusterB}{\ensuremath{\mathcal{G}^{\mathrm{B}}}} 
\newcommand{\NCB}{\ensuremath{N^{\mathrm{B}}}} 

\newcommand{\NCC}[2]{\ensuremath{N^{\mathrm{C}}_{{#1 #2}}}}
\newcommand{\clusterBelem}[1]{\ensuremath{B_{{#1}}}}
\newcommand{\clusterS}[1]{\ensuremath{\mathcal{G}^{\mathrm{S}}_{{#1}}}}
\newcommand{\NCS}[1]{\ensuremath{N^{\mathrm{S}}_{{#1}}}}
\newcommand{\clusterSelem}[2]{\ensuremath{S_{{#1 #2}}}}
\newcommand{\clusterC}[2]{\ensuremath{\mathcal{G}^{\mathrm{C}}_{{#1 #2}}}}
 
\newcommand{\clusterCelem}[3]{\ensuremath{C_{\mathcal{D}, {#1 #2 #3}}}}

\newcommand{\modelf}[1]{\ensuremath{\mathbb{#1}}}

\newcommand{\intscat}{\ensuremath{\sigma_{\mathrm{int}}}}

\begin{document}



\title{Neutron star mass and radius measurements from atmospheric model fits to X-ray burst cooling tail spectra}

\titlerunning{Atmosphere model fits of X-ray burst spectra}

\author{J.\,N\"attil\"a\inst{1,2}
  \and M.\,C.\,Miller\inst{3}
  \and A.\,W.\,Steiner\inst{4}
  \and J.\,J.\,E.\,Kajava\inst{5,1,6}
  \and V.\,F.\,Suleimanov\inst{7,8}
  \and J.\,Poutanen\inst{1,2,9}
}

\institute{Tuorla Observatory, Department of Physics and Astronomy, University of Turku, V\"ais\"al\"antie 20, FI-21500 Piikki\"o, Finland\\ \email{joonas.a.nattila@utu.fi}
  \and Nordita, KTH Royal Institute of Technology and Stockholm University, Roslagstullsbacken 23, SE-10691 Stockholm, Sweden
  \and Department of Astronomy and Joint Space-Science Institute, University of Maryland, College Park, MD 20742-2421, USA
  \and Department of Physics and Astronomy, University of Tennessee, Knoxville, Tennessee 37996, USA
  \and Finnish Centre for Astronomy with ESO (FINCA), University of Turku, V\"{a}is\"{a}l\"{a}ntie 20, FIN-21500 Piikki\"{o}, Finland
  \and European Space Astronomy Centre (ESA/ESAC), Science Operations Department, 28691 Villanueva de la Ca\~nada, Madrid, Spain
  \and Institut f\"ur Astronomie und Astrophysik, Kepler Centre for Astro and Particle Physics, Universit\"at T\"ubingen, Sand 1, D-72076 T\"ubingen, Germany
  \and Astronomy Department, Kazan (Volga region) Federal University, Kremlyovskaya str. 18, 420008 Kazan, Russia
  \and Kavli Institute for Theoretical Physics, University of California, Santa Barbara, CA 93106, USA 
}

\date{Received XXX / Accepted XXX}

\abstract{
Observations of thermonuclear X-ray bursts from accreting neutron stars (NSs) in low-mass X-ray binary systems can be used to constrain NS masses and radii. 
Most previous work of this type has set these constraints using Planck function fits as a proxy: both the models and the data are fit with diluted blackbody functions to yield normalizations and temperatures which are then compared against each other. 
Here, for the first time, we fit atmosphere models of X-ray bursting NSs directly to the observed spectra. 
We present a hierarchical Bayesian fitting framework that uses state-of-the-art X-ray bursting NS atmosphere models with realistic opacities and relativistic exact Compton scattering kernels as a model for the surface emission.  
We test our approach against synthetic data, and find that for data that are well-described by our model we can obtain robust radius, mass, distance, and composition measurements. 
We then apply our technique to {\it Rossi X-ray Timing Explorer} observations of five hard-state X-ray bursts from 4U~1702$-$429. 
Our joint fit to all five bursts shows that the theoretical atmosphere models describe the data well but there are still some unmodeled features in the spectrum corresponding to a relative error of 1--5\% of the energy flux. 
After marginalizing over this intrinsic scatter, we find that at 68\% credibility the circumferential radius of the NS in 4U~1702$-$429 is $R = 12.4\pm 0.4~\mathrm{km}$, the gravitational mass is $M=1.9\pm 0.3\Msun$, the distance is $5.1 < D/\mathrm{kpc} < 6.2$, and the hydrogen mass fraction is $X < 0.09$.
}


\maketitle


\section{Introduction}\label{sec:intro}

The masses and radii of neutron stars (NSs) encode valuable information about the properties of the matter in their cores \citep{Lattimer12ARNPS, LS14b}, which reaches several times nuclear saturation density and has strong isospin asymmetry, and which therefore cannot be analyzed in terrestrial laboratories.
Hence, detailed measurements of NS masses and radii are invaluable in the study of cold dense matter, and in particular the equation of state (EoS) of the matter, i.e., the relation between thermodynamic quantities such as the pressure and the energy density (see \citealt{Miller13, Ozel13, ML16} for recent discussions).  
The reliability of such constraints depends on the degree to which systematic errors can be controlled (in many current analyses such errors are significantly larger than the formal statistical uncertainties; see \citealt{Miller13,ML16}), as well as on the precision of the astrophysical models that are applied to the data.

One type of source that has attracted considerable attention in this context is low-mass X-ray binaries (LMXBs) that exhibit frequent thermonuclear X-ray bursts \citep[for reviews, see][]{LvPT93, SB06}.  By collecting observations from these bursts and modeling how they cool down, we can set constraints on the size of the emitting area (for early work in this field, see, e.g., \citealt{Ebi87,Damen90,vP90, LvPT93}).  It was, however, only the \textit{Rossi X-ray Timing Explorer} (\textit{RXTE}) that was able to produce a large catalogue of observations to study \citep[see, e.g.,][]{GMH08}.  Since then, a large number of bursts from different sources have been put to use \citep[for recent reviews, see][]{ML16, SPK16}.  There are currently two principal methods that are being used to infer the gravitational mass $M$ and the circumferential radius $R$ from burst cooling tails, both of which stem from the earlier work: the touchdown method (e.g., \citealt{Ozel06, OGP09, GO10, OPG16}) and the cooling tail method (e.g., \citealt{SPRW11, PNK14, NSK16,SPN17}).  Both methods fit the observed emission using Planck function and then compare the evolution of the observed temperature and normalization to the predictions of models \citep[see, however,][for an early attempt to circumvent the usage of Planck function fits only]{KMK11}.  
These fits simplify model comparison significantly, because the observed spectra are relatively well described by thermal emission.
However, because neither model atmosphere spectra \citep[e.g.][]{SPW12} nor the most accurately measured observed spectra \citep{MBL11} are exactly Planckian, using Planck fits as proxies throws away information and could even introduce biases.

Here we present, for the first time, simultaneous direct atmosphere model fits to a set of X-ray burst observations.  
We begin by studying the constraints that can be obtained from synthetic data, for which our model is a good description.  
This allows us to assess the accuracy of our method and to explore  possible biases in the results.  
We then apply our method to data from five hard-state bursts of 4U~1702$-$429.
We obtain interesting constraints on the mass and radius of this star and also study some of the previously neglected physical assumptions present in the fitting procedures.  
The LMXB system 4U~1702$-$429 is a particularly good testbed for the fitting as there has already been cooling tail modeling of the five hard-state bursts from this source \citep{NSK16}. 
Our initial analysis suggests that direct fitting of detailed atmosphere models to data is a promising avenue for extracting neutron star masses and radii from X-ray burst data.

In Sect.~\ref{sect:methods} we present the theoretical basis for our analysis.
This section is split into two parts: in Sect.~\ref{sect:model}, we describe how to model the emergent radiation and to couple it to the actual observations, and in Sect. \ref{sect:bayes}, we formulate the Bayesian framework and present our hierarchical fitting model.  
We apply our model to synthetic data in Sect.~\ref{sect:synt} and then to real X-ray burst observations from 4U 1702$-$429 in Sect.~\ref{sect:1702}.  
In Sect.~\ref{sect:1702} we also present our new improved mass, radius, distance, and composition constraints for the source.  
In Sect.~\ref{sect:disc} we discuss our results.  Finally, we present our summary in Sect.~\ref{sect:summary}.

\section{Methods}\label{sect:methods}

\subsection{Model for the emerging radiation}\label{sect:model}

Suppose that radiation from a point on the surface of a NS is initially emitted with a local specific intensity $I'_{E'}$ at energy $E'$, as measured in the local frame of the emission.  Assuming that the radiation propagates through vacuum to a distant observer, that observer will detect this radiation at energy $E$, where the energies are related by
\be
\frac{E}{E'} = \frac{1}{1+z},
\ee
where $z$ takes into account both the rotation-induced Doppler shifts and the gravitational redshift.  In the limit of low spin frequency,  $\nu \rightarrow 0$, the external spacetime is Schwarzschild and there are no Doppler shifts, and therefore the net redshift approaches
\be\label{eq:redshift}
\lim_{\nu \rightarrow 0} (1+z) = \left(1 - \frac{2 G M}{c^2 R} \right)^{-1/2},
\ee
where $G$ is the gravitational constant, $c$ is the speed of light, and $M$ and $R$ are respectively the gravitational mass and the circumferential radius of the NS.  A distant observer will measure a specific (monochromatic) intensity $I_E$ that is related to the original specific intensity $I'_{E'}$ \citep[Liouville's theorem for photons, see, e.g.,][]{MTW73, RL79} by
\be
I_{E} = \left( \frac{E}{E'} \right)^3 I'_{E'}.
\ee
The total observed monochromatic flux from the star, as seen by a distant observer, is then \citep[see, e.g.,][]{NP17}
\be\label{eq:fluxint}
F_{\mathrm{obs}}(E) = \int I_E d\Omega ,
\ee
where $d\Omega$ is the solid angle that the surface element occupies on the observer's sky.
In this paper we consider a uniformly emitting, slowly rotating NS.  In this limit, the observed flux is related in a simple way to the flux $F'$ emitted at the NS surface:
\be \label{eq:FobsE}
F_{\mathrm{obs}}(E) =
\frac{F'(E')}{(1+z)^3}  \left( \frac{R_{\infty}}{D} \right)^2 = \frac{F'(E')}{1+z}  \left( \frac{R}{D} \right)^2 ,
\ee
where $R_{\infty} = R(1+z)$ is the apparent NS radius, $D$ is the distance to the source, and $F'(E')=2\pi \int_0^1 I'_{E'} (\mu)\mu d\mu $ where $\mu$ is the cosine of the angle between the local normal direction and the direction of emission of radiation.

In general, a burster is not expected to emit uniformly, and rotation rates of known bursters extend up to 620~Hz \citep{MC02, Watts12}.  
Rotation introduces Doppler shifts that vary over the surface of the star and therefore smear sharp spectral features such as line.
These shifts also broaden continuum spectra, but such broadening can usually be neglected \citep{NP17}.
Moreover, the assumption of uniform emission combined with slow rotation means that the observed flux depends on  the surface flux and distance, but not the angular dependence of the specific intensity.  
This is not true in more general situations.  
We employ these approximations because they allow us to  simplify the general equation \eqref{eq:fluxint} and avoid the usage of computationally costly ray tracing to combine the flux from different parts of the star.  They also allow us to neglect a few potentially important but often unknown parameters such as the NS rotation frequency and the observer's inclination angle as well as the unknown latitude dependence of the flux.  

The gravitational acceleration at the NS surface is given by
\be \label{eq:gravity}
g = \frac{G M}{R^2} (1+z) \; .
\ee
The shape of the emerging spectrum has a weak dependence on $g$.
The composition of the atmosphere also affects the spectrum via the energy dependence of the opacity $\kappa$, which includes contributions from both true absorption and scattering.  For example, for an atmosphere with a hydrogen mass fraction $X$, the Thomson scattering opacity is
\be
\kappa_{\mathrm{T}} \approx 0.2(1 + X)~\mathrm{cm}^2\,\mathrm{g}^{-1}.
\ee
Assuming the Thomson opacity and a spherically symmetric flux, the outward radiative acceleration balances the inward gravitational acceleration at the stellar surface when the stellar luminosity reaches the Eddington luminosity $L_{\mathrm{Edd}}$, which is defined by
\be \label{eq:Ledd}
L_{\mathrm{Edd}} = \frac{4\pi G M c}{\kappa_{\mathrm{T}}} (1+z).
\ee
The actual critical luminosity is reached when the radiative acceleration 
\be
g_{\mathrm{rad}} = 
\frac{\kappa_{\rm R}}{c} \mathcal{F} 
\ee
equals the surface gravitational acceleration $g$.
Here $\kappa_{\rm R}$ is the flux mean opacity (equal in our case to the Rosseland mean opacity), 
\be
\mathcal{F} = \int F'(E') dE' = \sigma_{\mathrm{SB}} T_{\mathrm{eff}}^4 
\ee
is the bolometric surface flux, $T_{\rm eff}$ is the effective temperature  of  radiation and   $\sigma_{\mathrm{SB}}$ is the Stefan-Boltzmann constant. 
At high temperatures close to the critical luminosity, the opacity is dominated by Compton scattering and is smaller than $\kappa_{\mathrm{T}}$ because of the Klein-Nishina effect \citep{Poutanen17}, resulting in a critical luminosity exceeding $L_{\mathrm{Edd}}$ by 5--10\% \citep{SPW12}.
We will use the ratio $\grg$ to measure the escaping flux from the star.  
   
Using Eq.~(\ref{eq:Ledd}) we can also introduce the bolometric Eddington flux
\be
\mathcal{F}_{\mathrm{Edd}} = \frac{L_{\mathrm{Edd}}}{4 \pi R^2} = \sigma_{\mathrm{SB}} T_{\mathrm{Edd}}^4 = \frac{g c}{\kappa_{\mathrm{T}}},
\ee
characterized by the corresponding Eddington temperature $T_{\mathrm{Edd}}$.
The corresponding observed Eddington flux then can be obtained integrating Eq.~(\ref{eq:FobsE}) over energies: 
\be \label{eq:Feddobs}
F_{\mathrm{Edd}} = 
A\ \frac{\mathcal{F}_{\mathrm{Edd}}}{(1+z)^4}  = 
A\ \sigma_{\mathrm{SB}} T_{\mathrm{Edd,\infty}}^4 ,  
\ee
where $A = (R_\infty/D)^2$ is related to the apparent angular size  of the star and $T_{\mathrm{Edd,\infty}}=T_{\mathrm{Edd}}/(1+z)$ is the redshifted Eddington temperature.

For the flux escaping from the NS surface $F'(E')$ we will use the spectra from the atmosphere models computed in \citet{SPW11,SPW12} and \citet{NSK15}.  These calculations are based on the stellar modeling program \textsc{atlas} \citep{K70, K93}, but modified to deal with high temperatures \citep{I03, SulP06, SW07} and to take into account Compton scattering \citep{SPW12} using an exact relativistic redistribution function \citep[see, e.g.,][]{PS96}.  The models are computed in hydrostatic equilibrium, using local thermodynamic equilibrium, and assuming a plane-parallel atmosphere structure.  Because of these approximations, our model is only valid when the atmospheric scale height is much less than the stellar radius, which means that $\grg$ must be less than and not too close to unity.

Here we limit our model spectra to the range $\grg = 0.2 - 0.98$, to avoid any physical complications occurring at low or high temperatures: at high temperatures the scale height is too large, and at low temperatures (relevant late in the burst tails) it is likely that ongoing accretion breaks the assumption that the observed radiation emerges only from the passively cooling neutron star surface.  
In practise, limiting $\grg$ to such a range implies focusing on the first $\sim 10$ seconds of the cooling tail.
The compositions computed by \citet{SPW12} were: pure hydrogen ($X=1$), pure helium ($X=0$, $Y=1$), and solar hydrogen-to-helium ratio $X=0.738$ with different metallicities ($0.01$, $0.1$, $0.3$, and $1.0$ of solar).  
Here, for simplicity, we consider the atmospheres with a metallicity that is $0.01$ of solar.  
Such a selection is possible and does not introduce a considerable error because metals will only affect the spectra at the very late stage of the burst, when the atmosphere has a sufficiently low effective temperature \citep{SPW11, SPW12}.
We, on the other hand, do not consider observations on such a late stage where this effect would play a role.
Note also that exact selection of the metallicity does not play a key role here because we do not consider cold atmosphere models (we use $\grg > 0.2$), for which photoionization edges start to dominate the spectral features. In addition, we consider surface gravities of $\log_{10} g = 14.0$, $14.3$ and $14.6$ (with $g$ is cgs units).

We use these models to obtain spectra with any given $\grg$, $\log g$, and $X$ by linearly interpolating (or in the case of $\log g$ also linearly extrapolating) the logarithm of the monochromatic fluxes on the model photon energy grid.  The model parameter limits are: $\grg = 0.2 - 0.98$, $\log g = 13.7 - 14.9$, and $X = 0 - 1$.  We checked the accuracy of our interpolations by comparing our results against actual model spectra that were computed between the original grid points, and found that the relative accuracy of the spectral energy flux was better than 1\%.

Of course, what we observe is not energy flux but rather photon counts in energy channels.  We therefore convert our model spectra to photon counts by convolving them with a response function $R(I, E)$ of a detector:
\be\label{eq:counts}
C_{\mathcal{M},i} = t_{\mathcal{D}} \int_0^{\infty} M(E) R(I, E)~dE,
\ee
where $M(E)$ is the photon number flux of the model at energy $E$ and $t_{\mathcal{D}}$ is the observing time.  Here the response function $R$ is proportional to the probability that an incoming photon of energy $E$ will be detected in channel $i$ and is a discrete function (i.e. a response matrix) such that
\be
R_{\mathcal{D}}(i,j) = \frac{\int_{E_{j-1}}^{E_j} R(i,E)~dE}{E_j - E_{j-1}},
\ee
for an energy range $E_{j-1}$ to $E_j$.  In addition, one must take into account that the data might have a non-zero background.  In this case we fit the observed background with some spectral model $F_{\mathrm{bkg}}(E)$, so that our total model photon flux at energy $E$ is
\be
M(E) = \frac{F_{\mathrm{bkg}}(E)}{E} + \frac{F_{\mathrm{obs}}(E)}{E}.
\ee

The background flux is often estimated by observations prior to or after the burst, but observational \citep{Yu99, vPL86, Kuulkers03, Chen11, intZand11, Serino12, Degenaar13, WGP13, Peille14, WGP15, DKC16, Koljonen16, Kajava17} and theoretical work \citep{Walker92,  Miller96, Ballantyne04, Ballantyne05} suggests that the burst can increase or decrease the background rate and even change its spectrum.
Thus background estimates from times near the burst are unreliable and use of them could introduce bias.  
This is one reason that we focus on bursts that occur during the hard spectral state: 
for such bursts, the persistent background emission is very weak ($\lesssim 1\%$ of the peak flux).  
Therefore, although in practice we estimate the background using a 16~s observation prior the burst, which we find is described well by blackbody plus power law components (\textsc{bbodyrad} and \textsc{powerlaw} in \textsc{xspec}), the background modeling is unimportant because the emission is dominated by the burst radiation.  
Results supporting this kind of static persistent emission treatment were also presented in \citet{Kajava17}, where it is concluded that even if the background emission varies during the burst, it is unlikely to contribute more than $1\%$ of the burst flux in the hard state.
In the soft state, on the other hand, one of the persistent emission components can brighten more than 10-fold during the bursts \citep{Kajava17c}.
Finally, we note that we multiply both the background and the theoretical burst spectra by an interstellar absorption model (similar to \textsc{phabs} in \textsc{xspec}) to account for the non-zero neutral hydrogen column depth.

\subsection{Hierarchical fitting  model for $M$ and $R$ constraints}\label{sect:bayes}

Next, we construct a framework for comparing the emission models to the actual observations of X-ray bursts.
In order to do this, we formulate a hierarchical model for a NS that has been observed to have  $\NCB$ bursts $\clusterBelem{k}$, where $k = 1,\ldots,\NCB$.
We denote the set of all bursts as $\clusterB \equiv \left\{ {\clusterBelem{k}} \right\}_{k=1}^{\NCB}$.
Each of these bursts \clusterBelem{k}\ has $\NCS{k}$ spectra, which we label as $\clusterSelem{j}{k}$, one for each time bin $j$.
The set of all spectra in \clusterBelem{k}\ is similarly denoted as $\clusterS{k} \equiv \left\{ {\clusterSelem{j}{k}} \right\}_{j=1}^{\NCS{k}}$.
Each spectrum \clusterSelem{j}{k} consists of a set $\clusterC{j}{k} \equiv \left\{ {\clusterCelem{i}{j}{k}} \right\}_{i=1}^{\NCC{j}{k}}$ with $\NCC{j}{k}$ measurements of counts \clusterCelem{i}{j}{k}, measured in the detector channel $i$.

For a single channel we can define the likelihood function as $L(\mathcal{M})_{i} = P(\mathcal{D} ~|~ \mathcal{M}, \mathcal{H})$ so that $P(\mathcal{D} ~|~ \mathcal{M}, \mathcal{H})$ is the probability of the data $\mathcal{D}$ given the model $\mathcal{M}$ and a set of assumptions $\mathcal{H}$.
When the source of experimental noise is due to the number of events arriving at the detector, the counting statistics are Poisson distributed.
Hence, to estimate the model's goodness-of-fit for one element $C_{\mathcal{D},i}={\clusterCelem{i}{j}{k}}$ in some arbitrary burst \clusterBelem{k} and spectrum \clusterSelem{j}{k}, we compute the likelihood for a Poisson distributed data as 
\be\label{eq:ind_prob}
L_{i} = \frac{(C_{\mathcal{M},i} )^{C_{\mathcal{D},i}} ~e^{-C_{\mathcal{M},i}} }{(C_{\mathcal{D},i})!}.
\ee
The joint likelihood for a single spectrum is then 
\be
L_{\mathrm{S}}(\mathcal{M}) = P(\left\{ {C_{\mathcal{D},i}} \right\}_{i=i_{\mathrm{min}}}^{i_{\mathrm{max}}} ~|~ \mathcal{M}, \mathcal{H}) = \prod_{i} L_{i},
\ee
where $i$ ranges from the first detector channel $i_{\mathrm{min}}$ to the last detector channel $i_{\mathrm{max}}$ used in the analysis.
Note that because in practice likelihoods can be extremely large or small, we instead use log likelihoods in our analysis.  The joint likelihood for a burst is $L_{\mathrm{B}} = \prod_{S} L_{\mathrm{S}}$, and the total joint likelihood for all bursts is $L = \prod_{B} L_{\mathrm{B}}$.

In the limit of high count rate the Poisson distribution is well approximated by a Gaussian distribution.
In this case the likelihood is proportional to $\exp(-\chi^2/2)$, where%
\footnote{
This is known as Pearson's weighting when the statistical error in the denominator is taken from the model.
Similarly, one could describe the error with the help of the data counts, as is done with the Neyman's weighting.
Both of them are biased (in the opposite directions) estimators of the model parameters \citep[see][]{Humphrey09}.
}
\be\label{eq:chi}
\chi^2 = \sum_{i} \frac{( C_{\mathcal{D},i} - C_{\mathcal{M},i} )^2}{C_{\mathcal{M},i} }.
\ee
Because $\chi^2$ is proportional to the log likelihood, the joint log likelihood for the spectra in a burst, and the joint log likelihood for all bursts, is the sum of the individual $\chi^2$ values.  We caution that, particularly in the higher-energy channels, it can be that there are not enough counts that the Poisson distribution is well-approximated by a Gaussian.  In this case $\chi^2$ is not a good approximation to the log likelihood, and this could contribute to the formally poor $\chi^2$ we find in Section~3.2 when we analyze data from 4U~1702$-$429.

It is also possible that we have underestimated the uncertainties in our data.  In this case it is typical, in a Bayesian analysis, to introduce an intrinsic scatter $\intscat$ into the system.
Physically the intrinsic scatter can be understood to originate either from the instrument calibration error or from the uncertainty in the actual model used, which as we recall is interpolated from tabulated points in the space of composition, surface gravity, and temperature.  The addition of intrinsic scatter is similar to the error expansions in frequentist methods where data errors are increased until the total $\chi^2/\mathrm{d.o.f.}$ is around unity.
The underlying idea is that intrinsic scatter acts to quantify and penalize our ignorance of the model: by increasing \intscat, the possible credible regions for other parameters also inflate to take into account that the data are not fully described by the model.
Mathematically this is done by convolving the original Gaussian distribution, $\mathcal{N}_{\sigma}(x)$ (where $\sigma$ can be taken to be $\sqrt{C_{\mathcal{M}} }$ in relation to the $\chi^2$ formulation), with another normal distribution with undefined error $\intscat$.  It is standard to assume, given no other knowledge, that the errors can be added in quadrature: $\sigma_{\rm tot}^2=\sigma^2+\intscat^2$.
The likelihood in Eq.~\eqref{eq:ind_prob} can then be replaced with
\begin{align}\begin{split}
    L_i &= \mathcal{N}_{\sigma} \ast \mathcal{N}_{\intscat}(C_{\mathcal{D},i} - C_{\mathcal{M},i} ) \\
    & = \frac{1}{\sqrt{2\pi(\sigma^2 + \intscat^2)}} \mathrm{exp}\left( - \frac{1}{2} \frac{(C_{\mathcal{D},i} - C_{\mathcal{M},i} )^2}{\sigma^2 + \intscat^2} \right),
\end{split}\end{align}
where $\ast$ marks the convolution of the two functions.
Taking the logarithm simplifies the latter expression to 

\be
\ln L_i = \mathrm{const}
- \frac{1}{2} \ln{(\sigma^2 + \intscat^2)} - \frac{1}{2} \frac{(C_{\mathcal{D},i}- C_{\mathcal{M},i})^2}{\sigma^2 + \intscat^2}.
\ee
From this we see that the expression reduces to $\chi^2$ when $\intscat \rightarrow 0$ and $\sigma = \sqrt{C_{\mathcal{M},i}}$, as the first two terms in the likelihood expression can then be ignored as constants.
It is important to notice that \intscat\ cannot be increased infinitely to get a better likelihood, because the $\ln$ term in the log-likelihood expression compensates for the last term.
Hence, there exists some balance between the two terms and \intscat\ can only grow to some finite value where the previously unexplained scatter in the observations is then explained by the model.

\begin{table*}[ht!] 
\begin{center}
\caption{Parameters of hierarchical fitting models.}
\label{tab:models}
\begin{tabular}{l c c c l}   
  \hline
  \noalign{\vskip 0.5ex}
   Model name & Global parameters  & Burst parameters & Spectrum parameters & Assumptions \\
              &                    & \clusterB       & \clusterS{k}        & $\mathcal{H}$ \\ 
  \noalign{\vskip 2ex}
  \hline
   Model \modelf{A}  & $M$, $R$, $D$      & --- & $\grg$                   & $X = 0$, $S_{\mathrm{f}} = 1$ \\
   Model \modelf{B}  & $M$, $R$, $D$, $X$ & --- & $\grg$                   & $S_{\mathrm{f}} = 1$ \\
   Model \modelf{C}  & $M$, $R$, $D$      & --- & $\grg$, $S_{\mathrm{f}}$ & $X=0$ \\
    Model \modelf{D}  & $M$, $R$, $D$, $X$, $\intscat$ & --- & $\grg$ & $S_{\mathrm{f}} = 1$ \\
  \hline
\end{tabular}
\end{center}
\end{table*}

The actual model spaces are then constructed hierarchically on top of different clusters of nested data groups.
Such a model with nested hierarchy can be physically motivated by considering our problem at hand:
We have a neutron star that has some model parameters that can characterize it (such as size, distance, $\dots$).
This neutron star will then exhibit bursts that could also have some model parameters (such as composition, ignition depth, $\dots$).
The bursts, however, all share the same parameters that the neutron star has and, hence, in the model parameter hierarchy the burst parameters appear lower.
The bursts, on the other hand, are constructed of a series of snapshots in time that we call energy spectra.
Again, one individual spectrum could have parameters dedicated only to that one particular spectrum or share some parameters among the other spectra in the burst.
Such a nesting of parameters we then call a hierarchical nested model in this paper.

In this work we consider four hierarchical models, \modelf{A}, \modelf{B}, \modelf{C}, and \modelf{D}, presented in Table \ref{tab:models}.
As an example, we next go through the models in more detail.
At the top level of model \modelf{A} we define 3 shared global parameters:
NS mass $M$, radius $R$, and distance $D$ to the star.
The combination of $M$ and $R$ then gives us the surface gravity $g$ and the redshift $1+z$. 
These can be combined with the distance to give the quantity $A$, which is proportional to the solid angle occupied by a NS on the sky. 
In addition to these basic parameters we can set a global composition of the accreted matter via the hydrogen mass fraction $X$, as is done in Model \modelf{B}.
The next level in the model involves the different bursts $\clusterBelem{k} \in \clusterB$ for which we do not introduce any cluster-specific sub-parameters in this work.
Going further in the hierarchy tree, each spectrum of the burst $\clusterSelem{j}{k} \in \clusterS{k} \in \clusterBelem{k}$ has always at least one individual parameter to sample: 
the effective temperature of the emerging spectrum as expressed through the parameter $\grg$.
In Model \modelf{C}, we also introduce the fraction $S_{\mathrm{f}}$ that emits (and we assume that the emitting portion emits uniformly).  $S_{\mathrm{f}}$ is a free parameter that enters the flux equation by modifying the apparent angular size $A' = S_{\mathrm{f}} A$.
As we will see later on, the real data are not fully described by the model.
To accommodate this deviation we, in the end, expand the model \modelf{B} by introducing a free intrinsic scatter to the system on the global scale.
We label this model \modelf{D}.
In contrast to models \modelf{A} to \modelf{C}, for \modelf{D} we choose a non-uniform phenomenological distance prior; this choice is motivated by the synthetic data results.
Intrinsic scatter is always, when present, sampled as $\lognat \intscat^2$, as it is a scale parameter in the model. 

When a parameter is not free, but has some fixed value, our model is said to have a set of physical assumptions $\mathcal{H}$ that implicitly enter the likelihood calculations.
The strictest set of assumptions $\mathcal{H}$ is imposed for model \modelf{A}, which assumes constant (uniform) emitting area $S_{\mathrm{f}} = 1$ and known non-varying chemical composition ($X = 0$ or $0.73$ in this work).
In Model \modelf{B}, the assumption of the chemical composition is relaxed. 
Similarly, in Model \modelf{C} we test the validity of the constant emitting area assumption.

On a purely theoretical basis, one would expect a model with every parameter defined as free in the lowest hierarchy level to be the least informative:
by allowing both the hydrogen fraction $X$ and the surface fraction $S_{\mathrm{f}}$ to evolve freely in time for each spectrum we could check the assumption of constant uniform emitting area and constancy of the chemical composition.
In practice, such freedom in the model is not, however, possible as $X$ and $S_{\mathrm{f}}$ are strongly correlated because they both affect the normalization of the flux.
While composition does have a slight effect on the shape of the spectral energy distribution, the current data do not allow any meaningful constraints without the additional normalization dependency.
These freedoms could be slightly limited by making the composition vary only from burst to burst. 
This would then allow us to study the time evolution of the composition on much longer timescales from burst to burst.
Another possibility would be to introduce a burst-specific $S_{\mathrm{f}}$ term into the model to allow variations between bursts, for example by a changing accretion disk inner radius.

\begin{figure}[!tb]
\centering
\includegraphics[width=9cm]{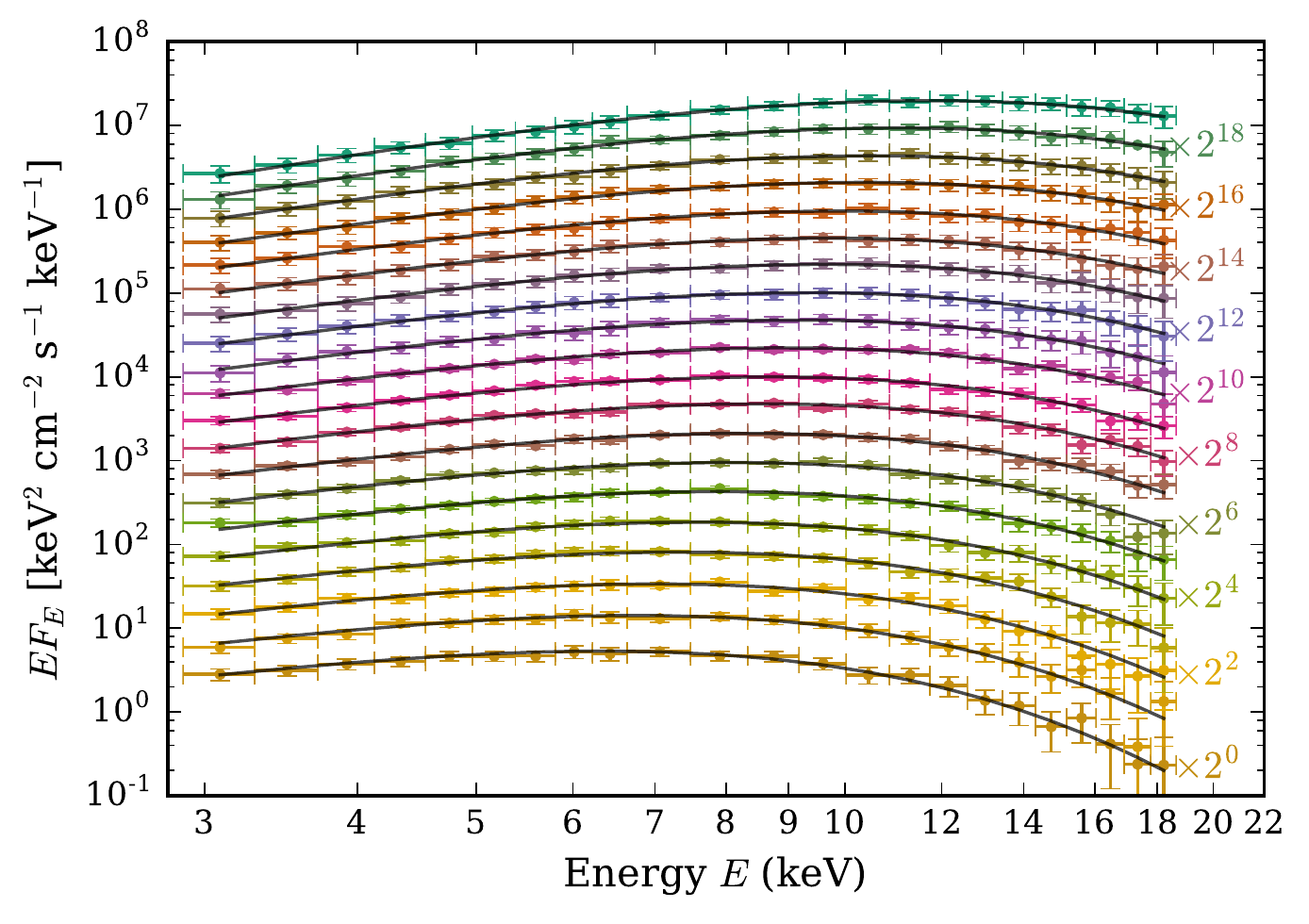}
\caption{\label{fig:synt_specs}
    Synthetic spectra (crosses) with corresponding best-fit atmosphere models (solid lines) for \modelf{A}.
Different colors show spectra for varying $\grg$.
Individual spectra are shifted by factors of $2$ in the y-direction for clarity. 
}
\end{figure}

\begin{figure}[!htb]
\centering
\includegraphics[trim={0 0.3cm 0 0},clip, width=9.0cm]{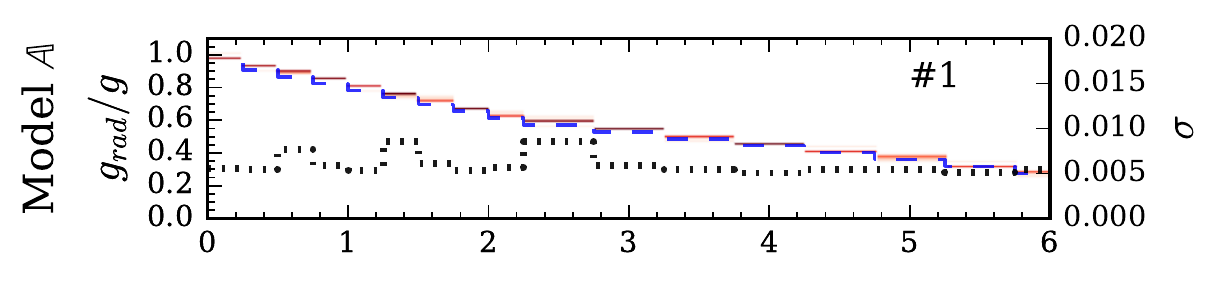}
\includegraphics[trim={0 0.3cm 0 0},clip, width=9.0cm]{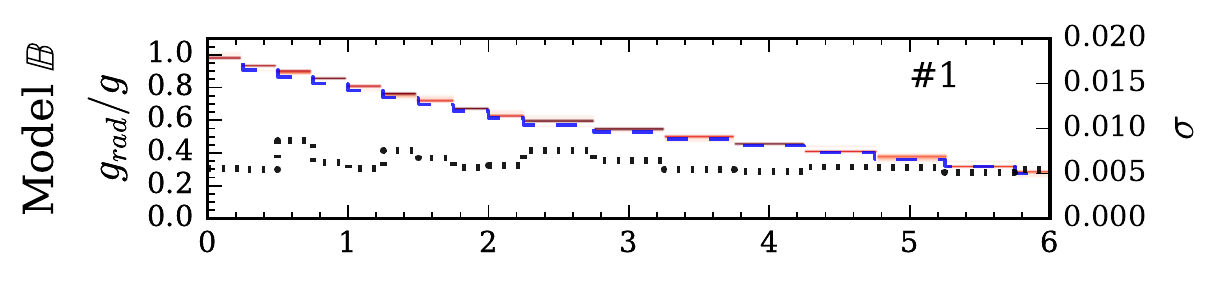}
\includegraphics[trim={0.1cm 0.3cm 0 0},clip, width=8.8cm]{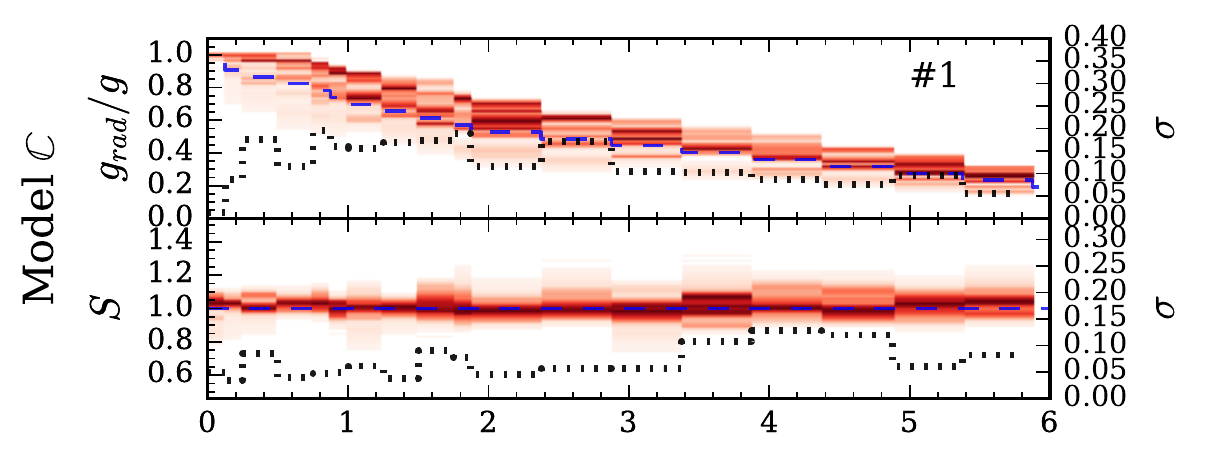}
\includegraphics[trim={0 0 0 0},clip, width=9.0cm]{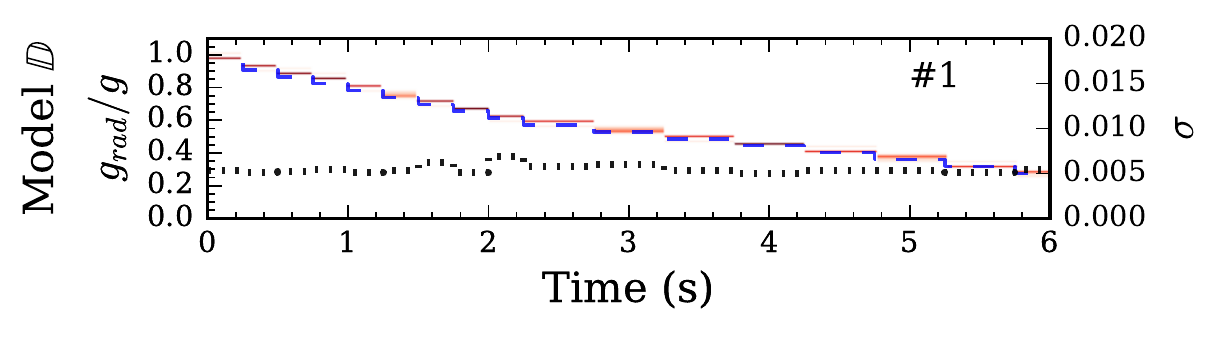}
\caption{\label{fig:synt_grg}
Evolution of the normalized luminosity $g_{\mathrm{rad}}/g$ for burst \#1 of the synthetic data.  Constraints on the surface fraction $S_{\mathrm{f}}$ are also shown for model \modelf{C}, which is the only one of our models in which this parameter is free.  The dotted black line with the scale on the right axis shows the corresponding standard deviation of the obtained parameter distributions.  
    The blue dashed line shows the input value used to create the data.  The darkness of the red coloring is proportional to the probability density.  This figure shows that when the fitting model is consistent with the model used to produce the data, we obtain parameter values that are accurate and precise. 
}
\end{figure}

Finally, we sample the parameter model space using Bayesian inference.
We introduce uniform prior distributions for $M$ and $R$ in the range $1.0 - 2.2~\Msun$ and $8-16~\mathrm{km}$, respectively.
For the distance, a uniform prior is taken in between $D = 2-10~\mathrm{kpc}$ 
For the model \modelf{D}, we choose to sample not $D$, but $D^{3/2}$, corresponding to a weakly informative prior of $\sqrt{D}$ for the distance, that slightly favors larger values.
Such a selection seems to remove the otherwise strong preference for smaller masses.%
\footnote{In our case, asserting an informative distance prior leads to a flat posterior in mass.
However, this will most likely also affect physical observables such as the flux.
Thus, even though we appear unbiased in mass, we are now biased in the flux.
Because of this, the aforementioned distance prior is only imposed for this one model to study the possible effects it might have.
}
We discuss this selection further in Sect.~\ref{sect:synt}.
When the hydrogen mass fraction is not fixed, we assume a flat prior ranging from $0$ (pure helium) to $1$ (pure hydrogen).
For the spectrum-specific nuisance parameters, we take similarly uniform limits so that $\grg = 0.2-0.98$ and $S_{\mathrm{f}} = 0.5-1.5$.
Note that values of $S_{\mathrm{f}} > 1$ are also allowed, as it might be possible that the apparent emitting area exceeds the one inferred from the star's angular size $A$ because of reflection from the accretion disk \citep{LS85}.
We also do \textit{not} impose any time relation in any of our models between neighboring time bins via adjacent $\grg$ values: every spectrum is free to attain any value in the prior range regardless of the adjacent spectra.
We then constrain the model parameters by sampling from the posterior distributions using Markov Chain Monte Carlo sampling.
To explore the parameter spaces we implement an affine-invariant ensemble sampler as discussed in \citealt{GW10} \citep[see also][]{GSW13}.
The actual implementation is heavily based on the \textsc{bamr} code \citep{bamr}, which on turn relies on the \textsc{o2scl} library \citep{o2scl}.
The ensemble sampler is similar to a normal Metropolis-Hastings algorithm but evolves not one but many parallel sample values called walkers together.
The random step for each walker is then done using a so-called stretch move algorithm where each walker makes a small step in the parameter space in relation to the whole ensemble.
Acceptance is still performed using a Metropolis-Hastings scheme.
With correlated distributions this will improve the autocorrelation times of the chains tremendously, allowing us to sample the parameter space more thoroughly in a shorter time.

\section{Analysis}\label{sect:analysis}

\subsection{Synthetic data}\label{sect:synt}

We begin our analysis by applying the methods described above to synthetic data.
The mock data is created to resemble the observations from NASA's \textit{RXTE} Proportional Counter Array (PCA) \citep{JMR06}.
We produce data using  $R=12~\mathrm{km}$, $M=1.5~\Msun$, $D = 6.0~\mathrm{kpc}$, and $X=0$.
These values are similar to those inferred from the five bursts of 4U 1702$-$429 that we analyze later \citep[see][]{NSK16}.
We also study similar NS configurations but with $X=0.73$, which thus corresponds to solar composition, to see how the composition affects the results.
The mock observations are created by computing the actual model spectra using the atmosphere models described in Sect.~\ref{sect:model}.
In this process, 20 spectra for each burst are created for 5 bursts in total so that $\grg$ is linearly spaced between values of $0.2$ and $0.95$.
For each spectrum we convolve the model with an actual \textit{RXTE}/PCA response matrix, compute the number of observed counts in each energy channel and then draw the observed number of counts from a Poisson distribution centered around the real value.
For the background spectra, we use the real background files from 4U 1702$-$429 hard state bursts. 
We increase the exposure time from $0.25$ seconds to $0.5$ seconds after 10 time bins; this procedure is commonly used for real data to keep the signal-to-noise level per time bin approximately constant despite the decreasing flux.
We fix the neutral hydrogen column depth to $N_{\mathrm{H}} = 1.87 \times 10^{22}~\mathrm{cm}^{-2}$, which again is similar to that of 4U 1702$-$429 \citep{WGP13}.
Fig.~\ref{fig:synt_specs} shows a set of spectra from one such synthetic burst.

\begin{figure*}
\centering
\includegraphics[width=9cm]{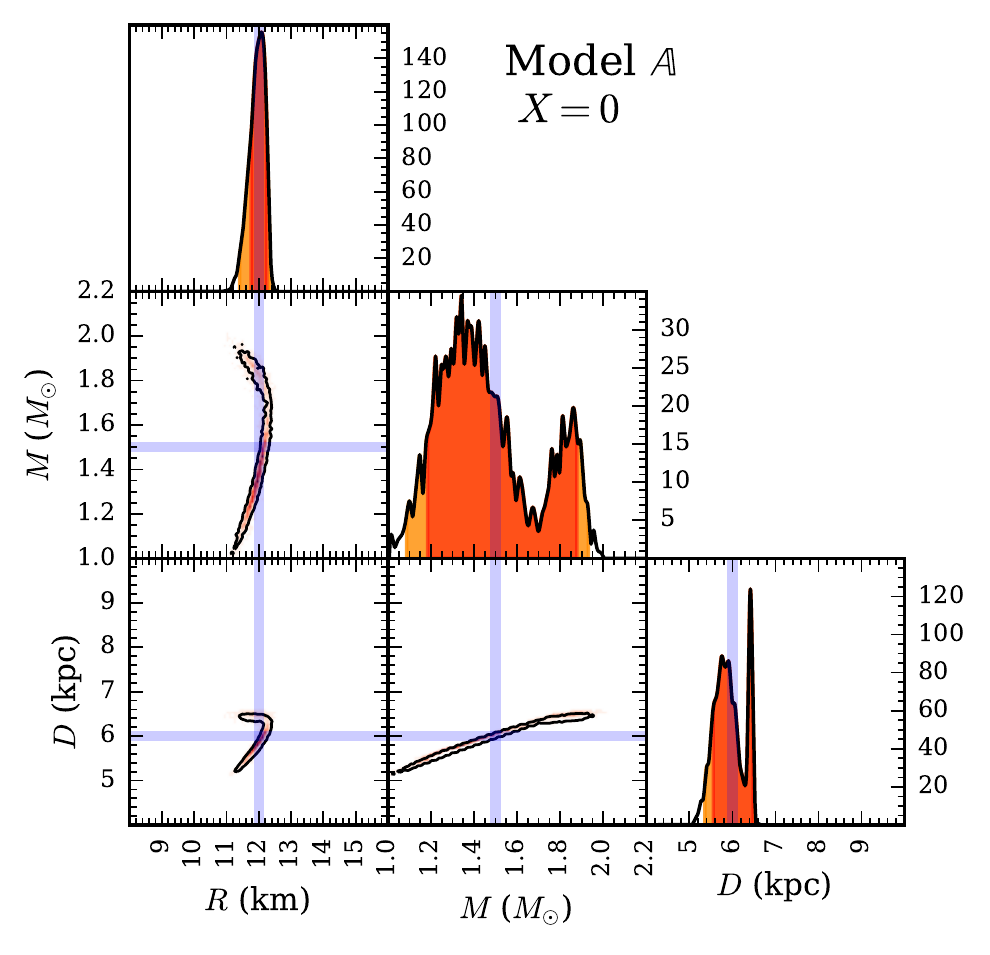}
\includegraphics[width=9cm]{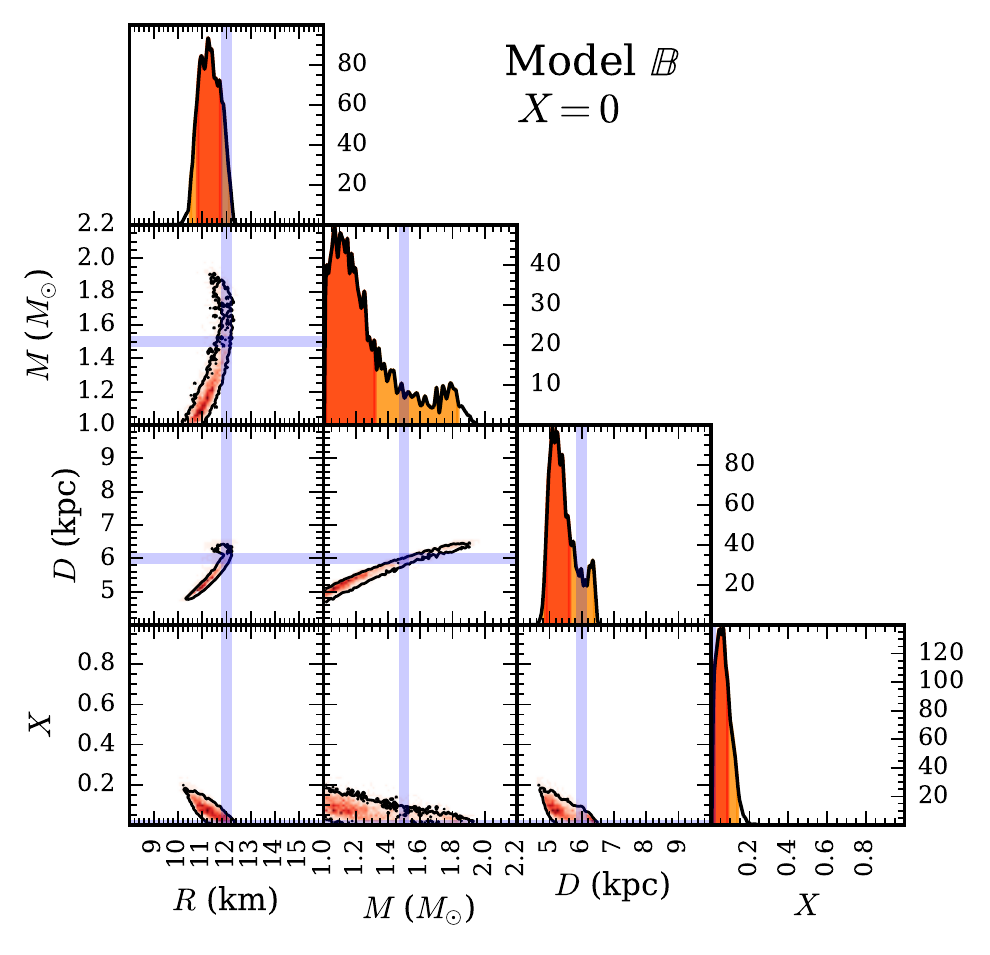}
\caption{\label{fig:synt_tri}
Posterior distributions for the MCMC run with synthetic data for Models \modelf{A} and \modelf{B}.
    The top panels of the triangles show the marginalized parameter posteriors (in arbitrary units); 
    here the dark and light orange shadings give the $68\%$ and $95\%$ credible regions.  
    The lower panels show the projected two-dimensional parameter posteriors against each other.  
    For these panels the solid line encloses the $95\%$ credible regions.  
    The blue stripes show the original values of the synthetic data, which were used to create the data: $R=12~\mathrm{km}$, $M=1.5~\Msun$, $X=0$, and $D=6~\mathrm{kpc}$.  
If the composition is known, the radius is precisely recovered.
When $X$ is a free parameter, the prior limit of $X > 0$ and the correlation of $M$, $X$, and $D$ with each other leads to asymmetric posteriors around the true value and so the inferred radius is a lower limit.
}
\end{figure*}

Next we fit this synthetic data with different models to assess how well we expect to constrain each model parameter.
This also gives us information about the possible biases in the method, given that we know the input values of the parameters. 
From here on, when we discuss credible regions we will always be referring to the highest posterior density regions.
First we study how well we can get information about nuisance parameters such as the normalized gravity $\grg$ and the fraction $S_{\mathrm{f}}$ of the surface that emits.
Fig.~\ref{fig:synt_grg} shows the evolution of these parameters.
Note also that our other parameters, such as $M$, $R$, and $D$, are allowed to vary freely.
When the surface fraction is fixed, we obtain the correct $\grg$ with a precision of about $0.02$ (in units of $g_{\mathrm{rad}}$), as can be seen from the width of the $68\%$ posterior distributions for models \modelf{A} and \modelf{B}.
If $S_{\mathrm{f}}$ is taken to be free (Model \modelf{C}), the uncertainty in $\grg$ increases by an order of magnitude, although the input value is within the uncertainty region.
$S_{\mathrm{f}}$ is determined with a precision of about $10\%$.
The lower precision for $S_{\mathrm{f}}$ is because only the spectral shape, rather than the amplitude, is used to match the model to the data.



\begin{figure*}
\centering
\includegraphics[width=8cm]{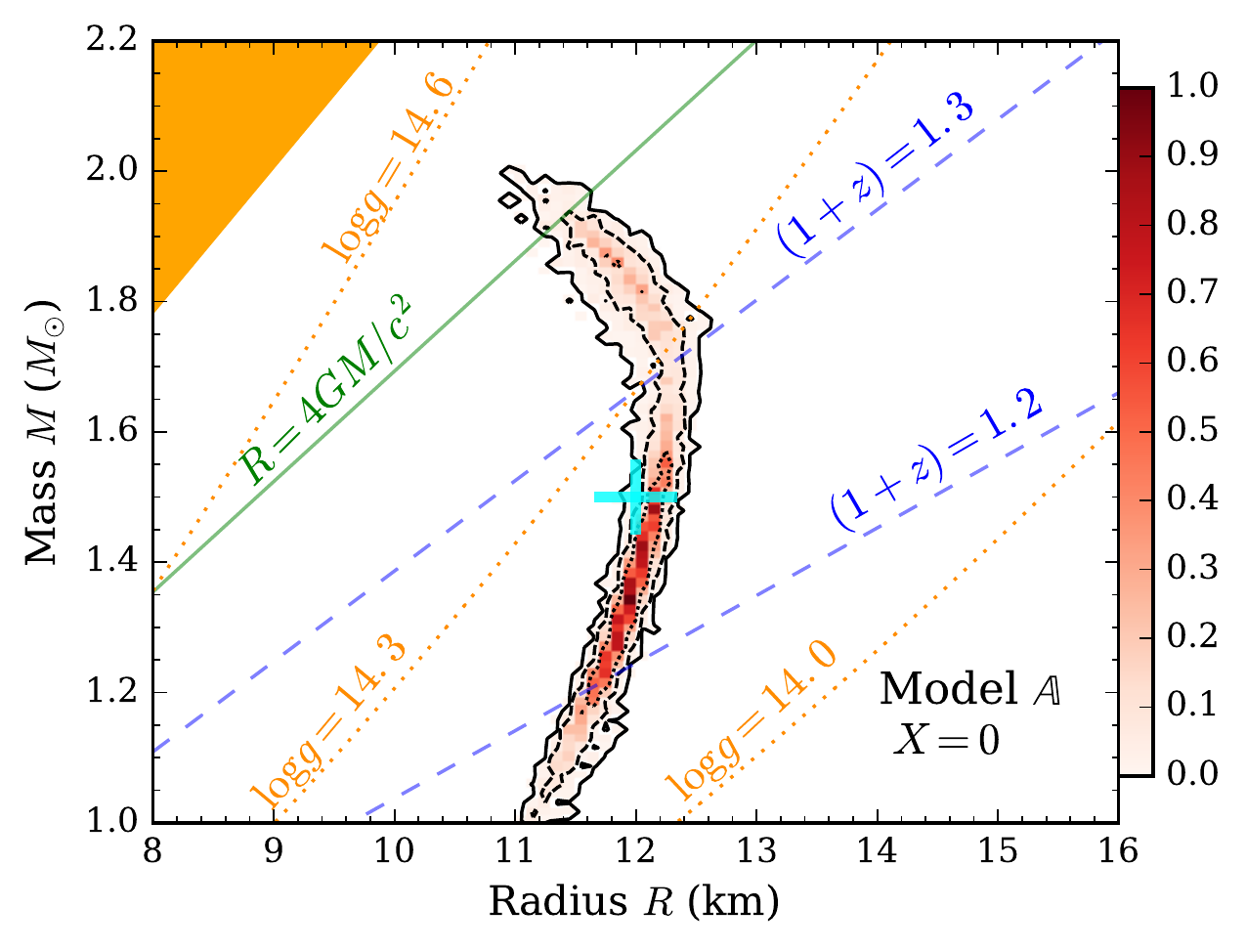}
\includegraphics[width=8cm]{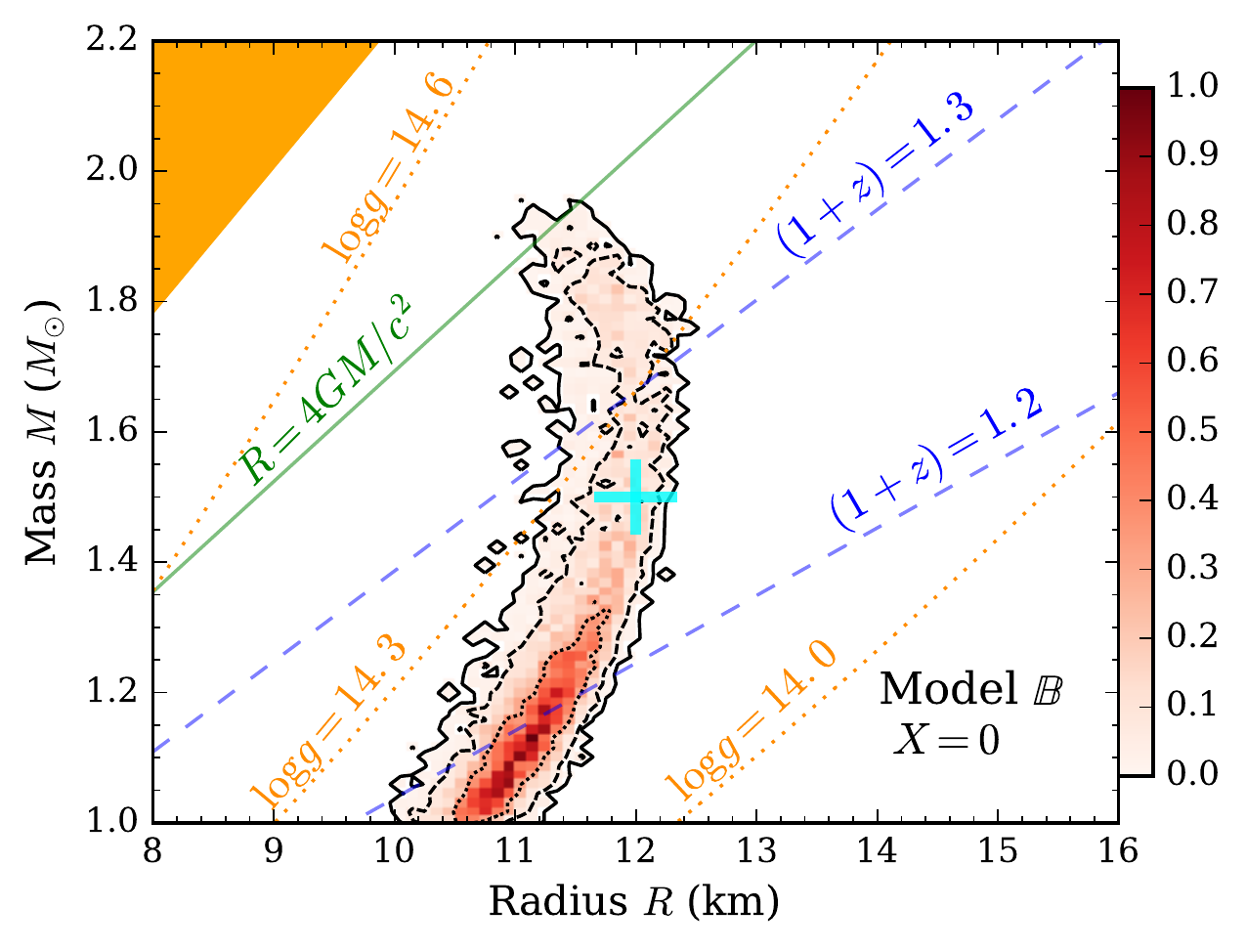}
\caption{\label{fig:synt_mr}
Mass and radius posteriors for synthetic data created for $R=12~\mathrm{km}$ and $M=1.5~\Msun$, which are shown here with cyan crosses.  The left panel shows a spectral fit with fixed emitting area $S_{\mathrm{f}}=1$ and hydrogen mass fraction $X=0$ (Model \modelf{A}).  The right panel shows a spectral fit with a free hydrogen fraction $X$ (Model \modelf{B}).  In both panels, the dotted line encapsulates the $68\%$, the dashed line the $95\%$, and the solid line the $99.7\%$  credible regions.  The dashed blue lines show values of constant redshift: $1+z=1.2$ and $1.3$.  The dotted orange lines are the contours of constant surface gravity: $\log g = 14.0$, $14.3$ and $14.6$. 
    The solid green line shows the critical radius $R=4GM/c^2$.
    The dark orange region at the top-left corner marks the region of parameter space forbidden by the requirement of causality \citep{HPY07, LP07}. 
    The input $(M,R)$ point is within the $95\%$ credible regions in both cases, even when the hydrogen mass fraction is a free parameter.
}
\end{figure*}


\begin{figure*}
\centering
\includegraphics[width=9cm]{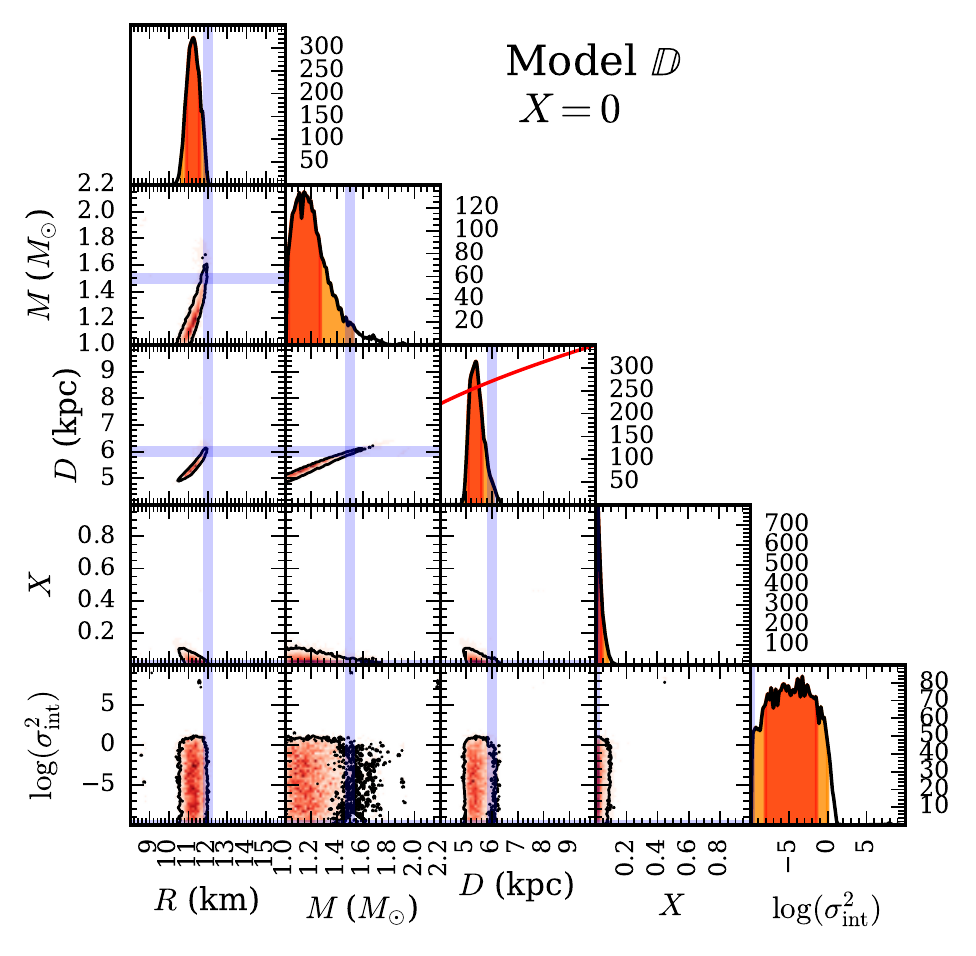}
\includegraphics[width=9cm]{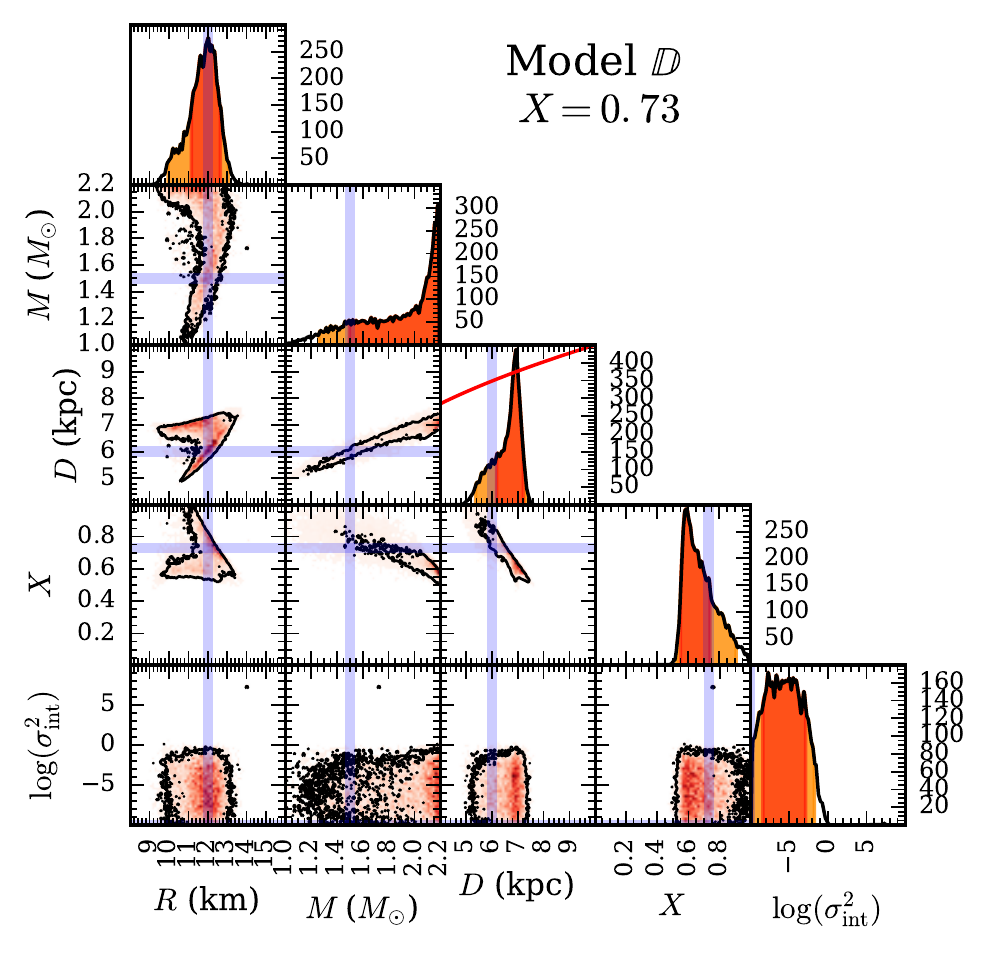}
\caption{\label{fig:synt07_tri}
    Posterior distributions for the MCMC run with synthetic data for the model \modelf{D} with helium and solar compositions.
    The red solid line in the $D$ panel shows the prior distribution ($\sqrt{D}$) that we used. 
    Other symbols and legends are the same as in Fig.~\ref{fig:synt_tri}. 
    With the inclusion of the weak distance prior, we see that the fit is better at recovering the original values.
    If the hydrogen mass fraction is not exactly zero, the true $R$, $M$, and $X$ are recovered when a correct family of solutions is considered.
    In many cases, the incorrect high-mass, small-radius family of solutions is easy to discard on a physical basis as it is close to or inside the region ruled out by causality.
}
\end{figure*}

\begin{figure*}
\centering
\includegraphics[width=8cm]{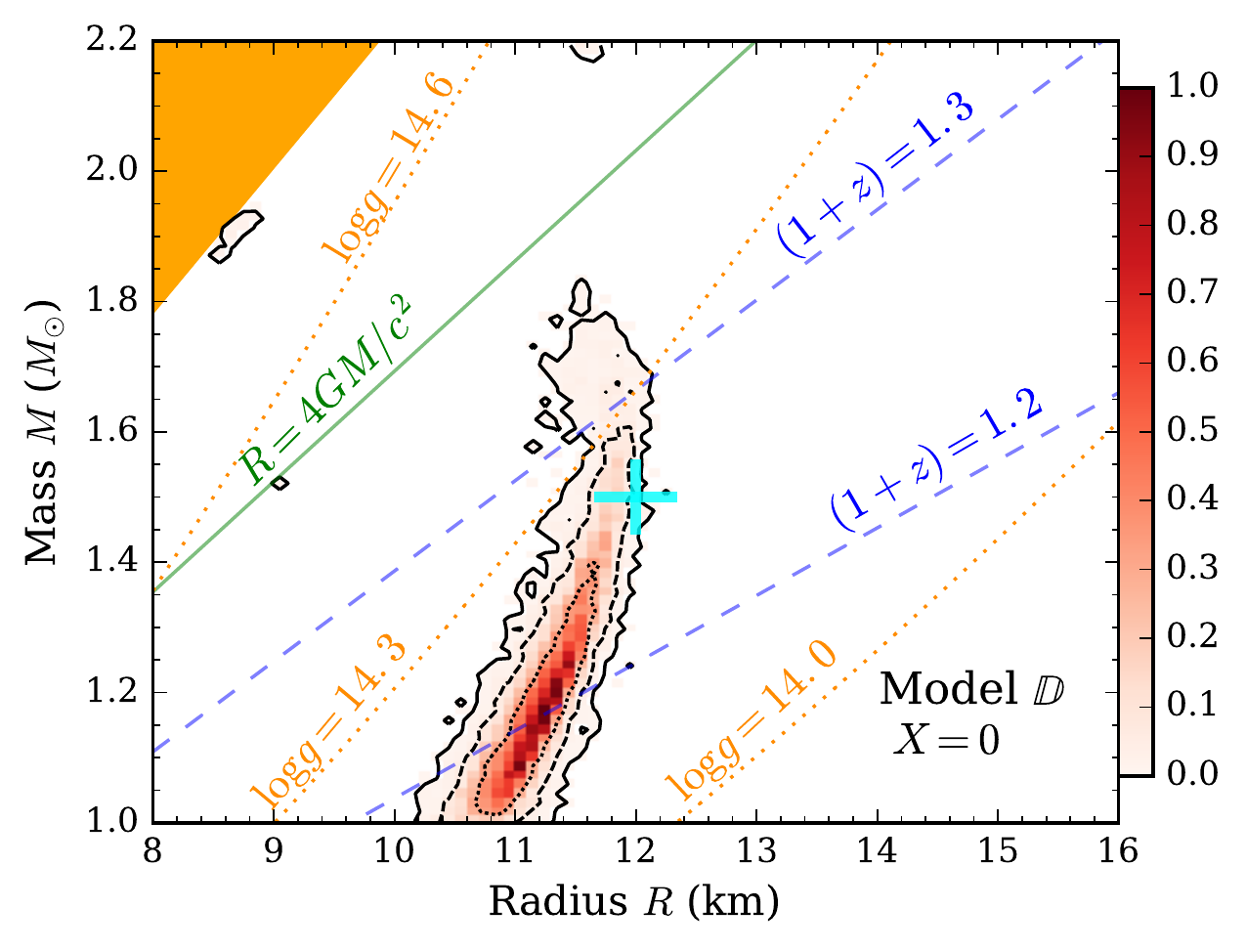}
\includegraphics[width=8cm]{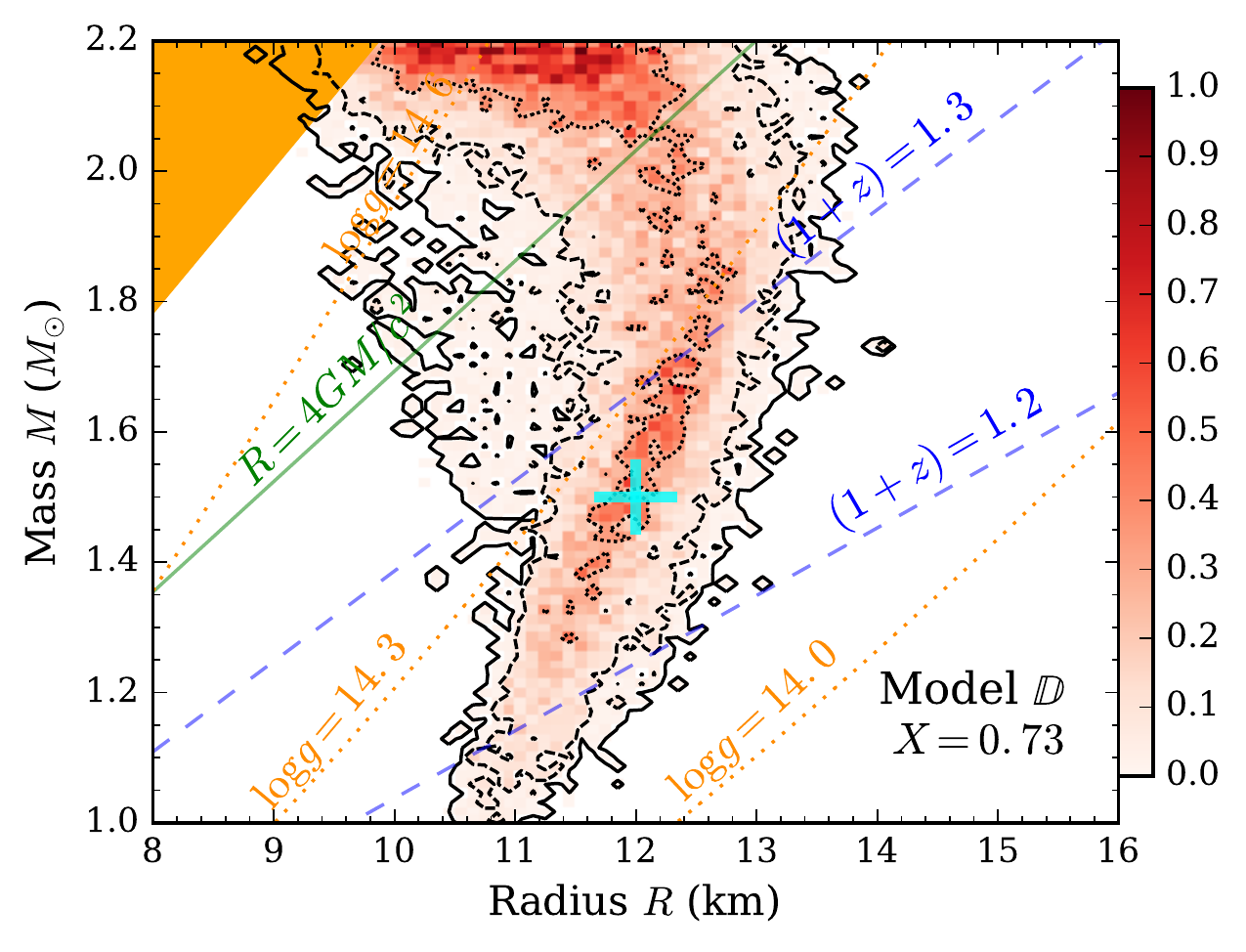}
\caption{\label{fig:synt07_mr}
Mass and radius posteriors for synthetic data created for $R=12~\mathrm{km}$ and $M=1.5~\Msun$, which are shown here with cyan crosses.
The left panel shows the results for a pure helium composition ($X=0$), whereas the right panel shows the results for a solar composition ($X=0.73$).
The symbols and legends are the same as in Fig.~\ref{fig:synt_mr}. 
    Even with a free hydrogen mass fraction $X$, the true $M-R$ values are now more consistent with the constraints coming from the fits. 
}
\end{figure*}

As a next step, we consider the parameters of greatest interest.
Fig.~\ref{fig:synt_tri} shows the marginalized parameter distributions for $M$, $R$, $D$, and $X$ (when sampled) along with the 2-dimensional posterior space projections.
Fig.~\ref{fig:synt_mr} shows the $M$-$R$ projection in more detail.
We see that  Model \modelf{A} is able to recover the correct radius $R=12~\mathrm{km}$ with an accuracy of about $0.7~\mathrm{km}$ after marginalization over all other parameters.
The mass, on the other hand, is always underestimated and shows a bimodal structure that is familiar from cooling tail fits \citep{SPRW11, PNK14, NSK16, SPN17}.
The input $(M,R)$ value is just at the boundary of the $95\%$ credible region.
We determine the distance with a $68\%$ scatter of about $0.6~\mathrm{kpc}$ and no bias.
The sharp spike in the marginalized distance distribution near the maximum value originates from the solutions near the critical line where $R=4GM/c^2$.
Mass and radius values on this line correspond to the solutions that are close to the maximum distance attainable for the system \citep[see e.g., Appendix A of][for more discussion]{PNK14}.

When the hydrogen fraction $X$ is a free parameter and the data are analyzed using Model \modelf{B} we see that both the radius and the mass are now underestimated.
Such an effect originates from the asymmetric $X$ distribution, which arises because the lower limit of $X=0$ is set by physical assumptions.
For larger $X$ we get smaller values of $M$ and $R$ than what we obtain using $X=0$.
In the $M-R$ plane our proper solution is now only inside the $95\%$ credible regions of the posteriors due to the strong bias toward smaller masses.
For similar reasons, the distance in this case is underestimated.
The hydrogen fraction is constrained to be $X < 0.2$ with $95\%$ credibility.

When $X=0.73$, there is a similar underestimation of the values of the parameters.
This is again caused by the connection between $X$, $D$, and $M$.
In this case, the posterior for $X$ is not symmetric around the true value because the distance has a maximum set by the observed flux level.
This causes $X$ to favor larger values (i.e., $X > 0.73$) and so the constraints on $M$ and $D$ are similar to the results from the model \modelf{B} analysis when $X=0$.

Ideally, we would like the method to be free from any bias in $M, R, D$, or $X$. 
These parameters are not, however, our observables. 
Quantities closer to the observations include the redshift given in Eq.~\eqref{eq:redshift}, which depends on $M$ and $R$, and the surface gravity $g$ defined by Eq.~\eqref{eq:gravity}, which is also a function of $M$ and $R$. 
The distance $D$ and the hydrogen fraction $X$ enter the system of equations via the flux (Eq.~\eqref{eq:FobsE}), which in turn is limited by the Eddington flux \eqref{eq:Feddobs}.  What we observe directly is the number of counts, which is related to the photon (number) flux of the source by Eq.~\eqref{eq:counts}.
Because of this, all of our parameters are interconnected in a complicated fashion.  In Bayesian inference, it is typical to study such a system by defining some information criterion and then to minimize its value given the fit parameters. 
This would then give us a description of the least informative, often multidimensional, priors.
We elect to instead impose simple uni-dimensional priors for the system based on experience with our analysis of synthetic data.
What we have found is that $M, X$, and $D$ are the most tightly connected parameters in the system.
Hence, imposing a prior distribution for one of them will strongly affect the rest.
In this work we have decided to optimize our results for $M$ and $X$ at the cost of introducing a non-flat prior for $D$.
Based on our different test runs we concluded that a prior of $P(D) \propto D^{1/2}$ (which therefore slightly favors larger values of $D$) produces the least biased constraints for $M$ and $X$.

This leads us to propose a fourth and final model, model \modelf{D}, which is an extension of model \modelf{B}.  
In model \modelf{D} the hydrogen fraction $X$ is a free parameter, but we also incorporate intrinsic scatter \intscat\ into the analysis and choose $P(D)\propto D^{1/2}$. 
Using these assumptions we recover the input parameters for the synthetic data, without any significant bias when $X \ne 0$.
This is evident in Figs.~\ref{fig:synt07_tri} and \ref{fig:synt07_mr}, where we use model \modelf{D} with synthetic data that have $X=0$ (left hand panels of both figures) and $X=0.73$ (right hand panels of both figures).
We find that when $X$ is not exactly $0$, we are able to reproduce the input radius without bias.
Additionally, if the second cluster of solutions at high masses is neglected, we also reproduce the input mass, hydrogen fraction, and distance.
In practice, this can be done by imposing a mass cutoff of $M<2.0~\Msun$ or by selecting only the low-mass solutions below the critical radius $R=4 G M/c^2$.
For a pure helium atmosphere, the sharp boundary at $X=0$ leads to asymmetric posteriors around the real value, and so the estimates are always biased towards smaller or larger values.
An additional check is that the intrinsic scatter is driven toward small values, which it must be because we created the data without any additional scatter.

For pure helium ($X=0$, Model \modelf{A}) we constrain the radius to be $R= 11.3 \pm 0.4 ~(0.7)~\mathrm{km}$ at $68\%$ ($95\%$) credibility.
The mass is similarly constrained to be $M= 1.2^{+ 0.2  ~(0.4)}_{-0.2  ~(0.2)}~\Msun$.
Thus the input values are inside the $95\%$ credible intervals.
The distance is found to be $D= 5.4^{+ 0.3 ~(0.7)}_{-0.4 ~(0.5)}~\mathrm{kpc}$.
The hydrogen mass fraction is constrained to be $X < 0.05 ~(0.09)$, which is consistent with the input value $X=0$. 
Similarly, for the synthetic solar data ($X=0.73$), the radius, distance, and hydrogen mass fraction results are 
$R= 12.0 ^{+ 0.7  ~( 1.1)}_{ -0.9  ~( 2.1)}~\mathrm{km}$,
$M= 2.2 ^{+ 0.1  ~( 0.1)}_{-0.7  ~( 1.0)}~\Msun$,
$D= 6.9^{+ 0.3 ~( 0.4)}_{-0.8 ~( 1.6)}~\mathrm{kpc}$, and
$X= 0.59^{+ 0.16  ~( 0.33)}_{-0.04  ~(0.06)}$.
Most importantly, we see that the correct radius is obtained without any bias.
If the second high-mass cluster of solutions is omitted by asserting an additional $M<2~\Msun$ prior, we obtain
$R= 12.0 ^{+ 0.6  ~( 1.0)}_{-0.9 ~(2.2)}~\mathrm{km}$,
$M= 1.5^{+ 0.4 ~( 0.4)}_{-0.3  ~(0.4)}~\Msun$,
$D= 6.9^{+ 0.4  ~(0.4)}_{-1.0  ~(1.6)}~\mathrm{kpc}$, and
$X= 0.59^{+ 0.15 ~( 0.31)}_{-0.05  ~( 0.07)}$.
In this case, both the radius and the mass are correctly recovered with $0.9~\mathrm{km}$ and $0.4~\Msun$ precision, respectively.
Even accounting for the mass imprecision when the composition is pure helium, model \modelf{D} is still the best of our models in producing precise and unbiased estimates of the parameters.

\subsection{4U 1702$-$429}\label{sect:1702}

\begin{figure}[!t]
\centering
\includegraphics[width=9cm]{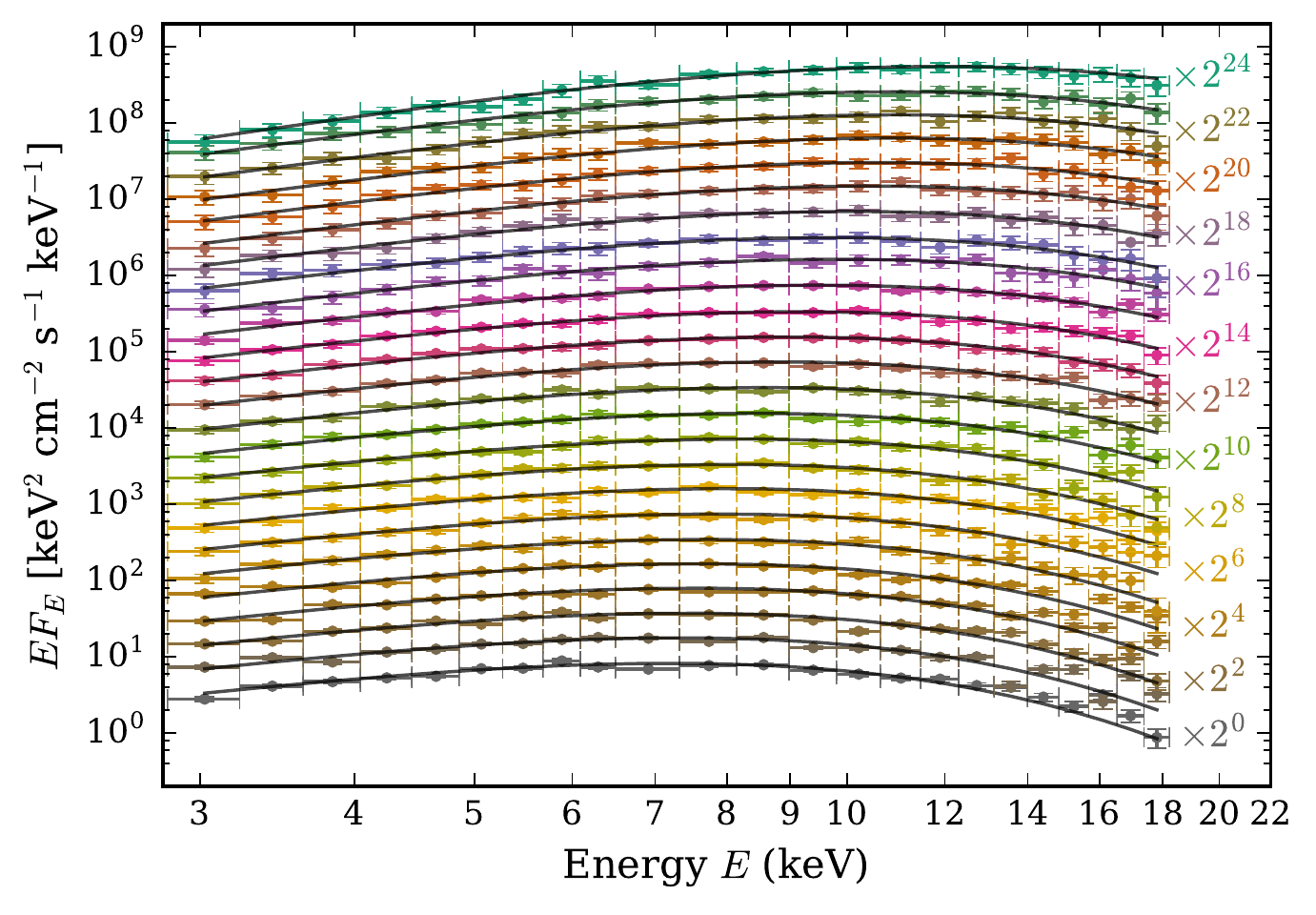}
\caption{\label{fig:data_specs}
Spectra for one X-ray burst from 4U 1702$-$429  (crosses) with corresponding best-fit atmosphere models (solid lines) for \modelf{A}.
Different colors show spectra for varying $\grg$.
Individual spectra are shifted in powers of $2$ in the y-direction for clarity.
}
\end{figure}

\begin{figure}
\centering
\includegraphics[trim={0 0.3cm 0 0},clip, width=9.0cm]{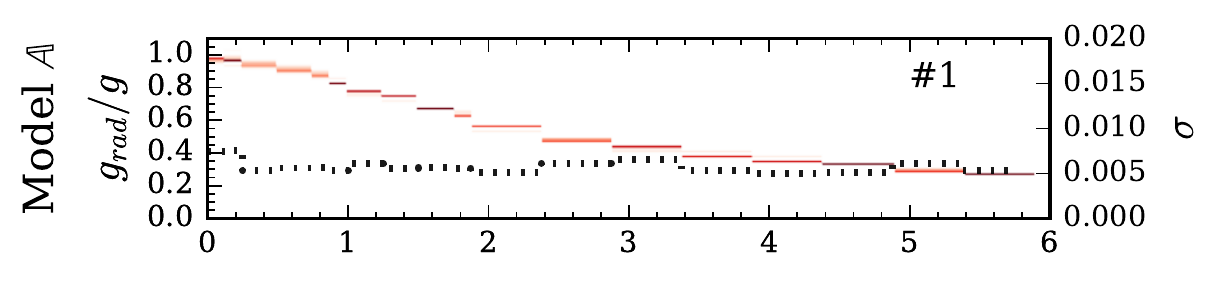}
\includegraphics[trim={0 0.3cm 0 0},clip, width=9.0cm]{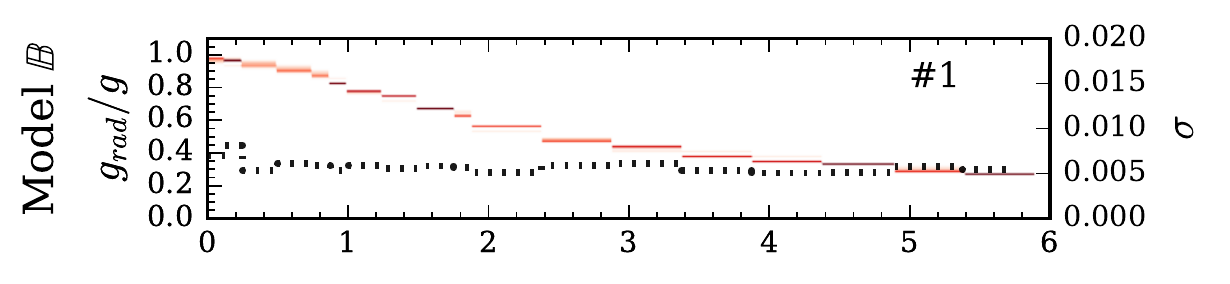}
\includegraphics[trim={0.1cm 0.3cm 0 0},clip, width=8.85cm]{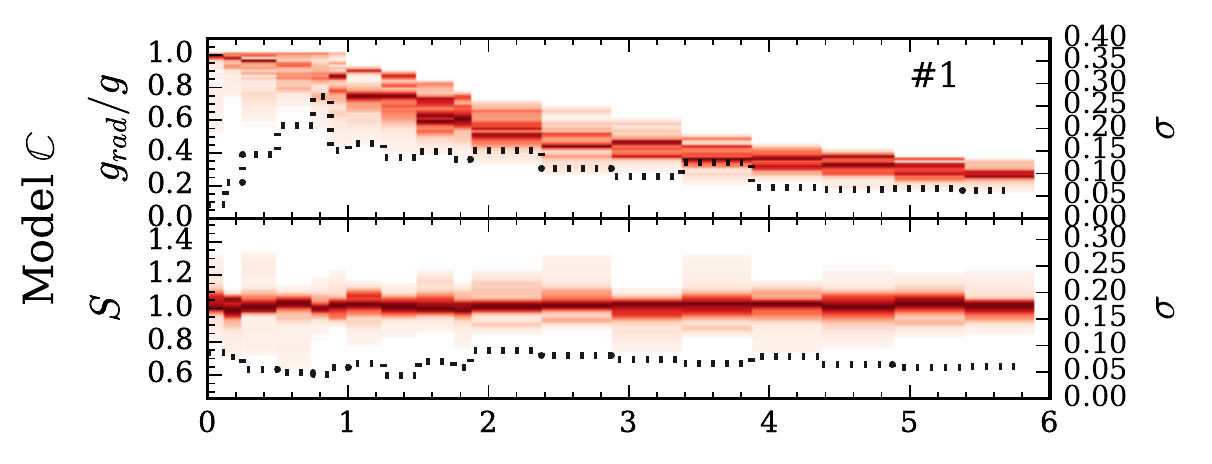}
\includegraphics[trim={0 0 0 0},clip, width=9.0cm]{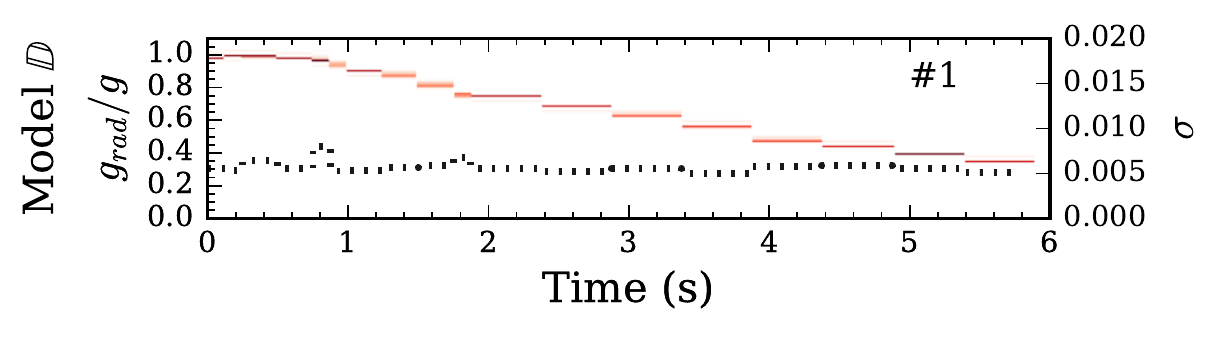}
\caption{\label{fig:1702_burst}
Evolution of the normalized luminosity $g_{\mathrm{rad}}/g$ for burst \#1 from 4U 1702$-$429.
The surface fraction $S_{\mathrm{f}}$ constraints are also shown for model \modelf{C}, which is the only one of our models in which this parameter is free.
The dotted black line with the scale on the right axis shows the corresponding standard deviation of the obtained parameter distributions.
The strength of the red coloring is proportional to the probability density.
    The constraints we find using this real {\it RXTE} data bear a clear similarity to the constraints from our synthetic data fits seen in Fig.~\ref{fig:synt_grg}. 
}
\end{figure}

We now study the {\it RXTE} data from 4U 1702$-$429.
The bursts we use are from obsid 50025-01-01-00, 80033-01-01-08, 80033-01-19-04, 80033-01-20-02, and 80033-01-21-00, starting at MJDs of 51781.333039, 52957.629763, 53211.964665, 53212.794286, and 53311.806086, respectively.
The data are reduced in a way similar to the reduction in \citet{GMH08} \citep[see also][]{KNL14, PNK14, NSK16}.  
As we described in Sect.~\ref{sect:synt}, we bin the data in time:
each time the count rate decreases by a factor of approximately $\sqrt{2}$, we double the exposure time so that the number of counts in each bin remains relatively high.  
The {\it RXTE} data were also deadtime corrected \citep[see, for example,][]{NSK16}, which was of course not necessary for the synthetic data.
One should also notice that unlike the synthetic data, the real observations have a varying ``quality'' due to the varying number of PCUs between the five bursts (ranging from 2 to 5 active PCUs).
Some sample spectra from one of the bursts are shown in Fig.~\ref{fig:data_specs}.  We also show the evolution of each individual spectral parameter of each burst in Fig.~\ref{fig:1702_burst}.  
Just as with the synthetic data, we see that the normalized luminosity $\grg$ is well constrained, and the evolution of this parameter is strikingly similar to its evolution in the mock data.  In the Model \modelf{C} fits we see that the surface emitting fraction is constrained to be very close to unity for the entirety of this burst, which provides some evidence that the full surface emits close to uniformly in this case.

\begin{figure*}[!htb]
\centering
\includegraphics[width=15cm]{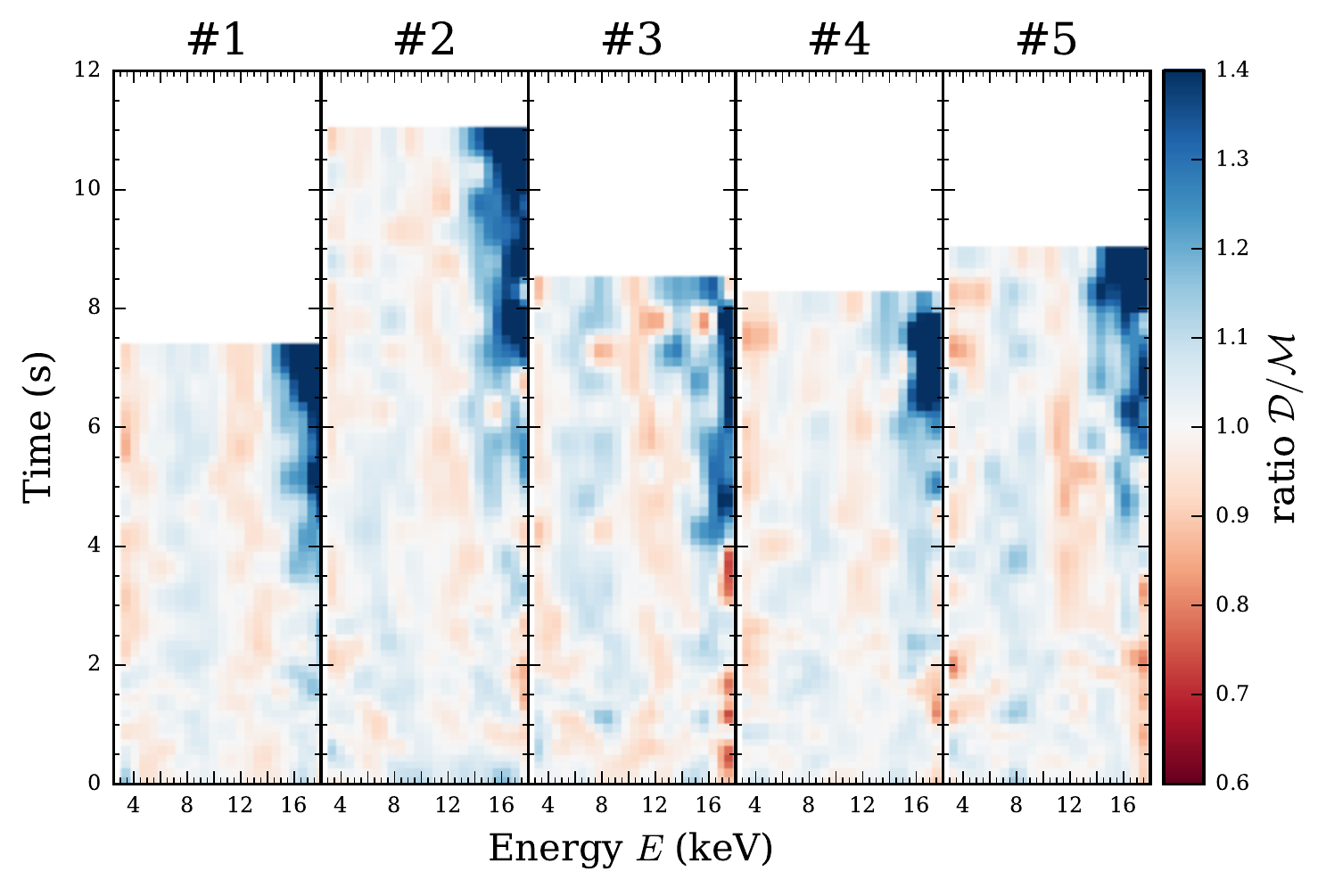}
\includegraphics[width=15cm]{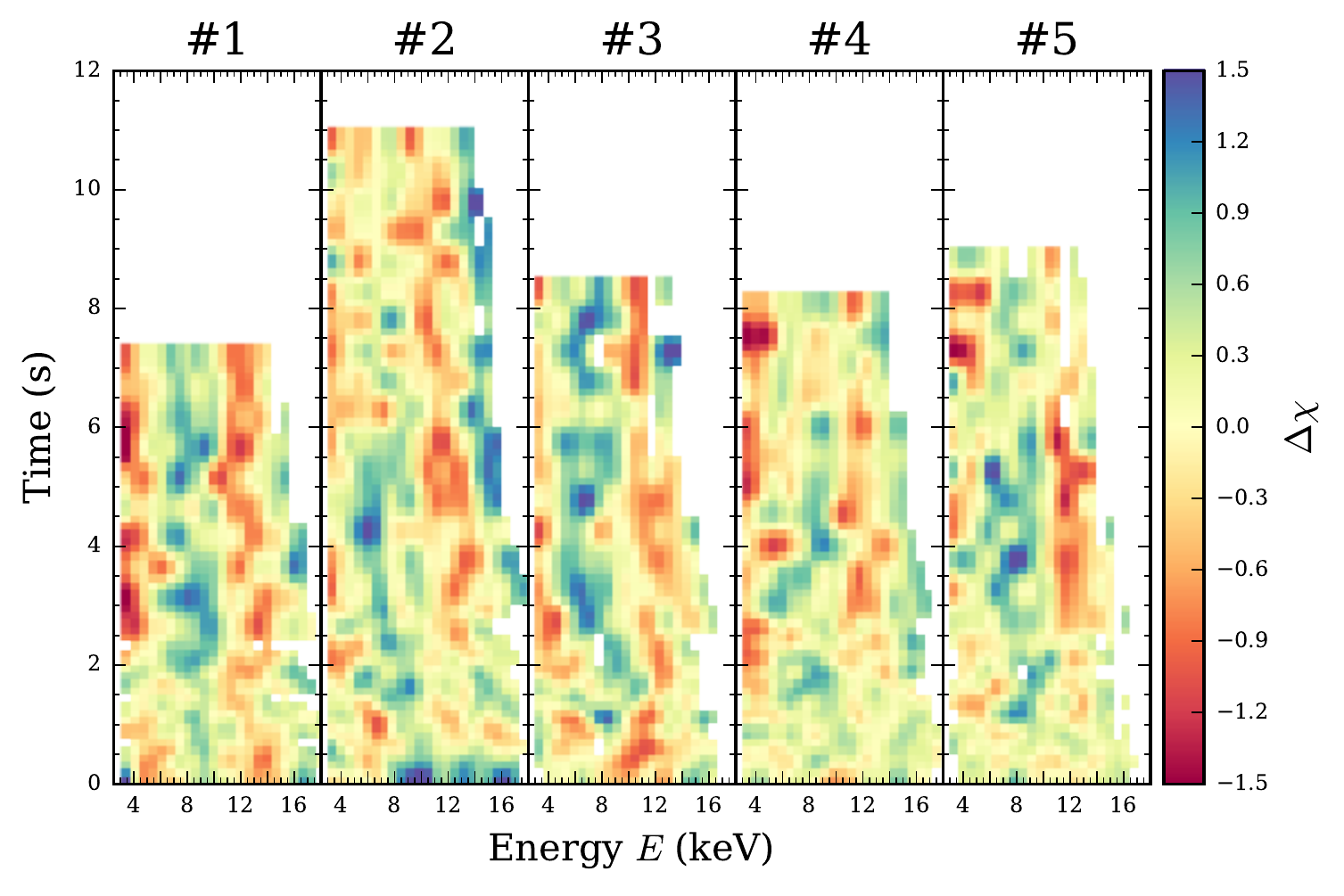}
\caption{\label{fig:ratio}
    {\it Upper panel}: Ratio of the data to the best-fit model for the time-resolved spectra of 5 bursts from 4U 1702$-$429 for Model  \modelf{D}.
{\it Lower panel:} Deviation  $\Delta\chi$ of the data  from the model. Only energy bins where the number of counts exceeds $50$ are shown.
}
\end{figure*}

\begin{table*}[ht!]
\begin{small}
\begin{center}
    \caption{The $\chi^2$ values for the atmosphere model best-fits for the Model \modelf{D}.} 
  \label{tab:chis50}
\begin{tabular}{l c c c c c c}
  \hline
  \noalign{\vskip 0.5ex}
   Bin  & Burst 1 & Burst 2 & Burst 3 & Burst 4 & Burst 5 & Synthetic burst 1\\
  \noalign{\vskip 2ex}
  \hline
1  & $ 25.4$ / $21$ ($1.21$) & $ 35.8$ / $21$ ($1.70$) & $ 19.6$ / $19$ ($1.03$) & $ 19.3$ / $21$ ($0.92$) & $ 17.7$ / $19$ ($0.93$)   & $ 27.3$ / $21$ ($1.30$)  \\
2  & $ 23.3$ / $22$ ($1.06$) & $ 27.5$ / $21$ ($1.31$) & $ 13.4$ / $20$ ($0.67$) & $ 17.8$ / $22$ ($0.81$) & $ 14.4$ / $20$ ($0.72$)  & $ 20.6$ / $21$ ($0.98$)   \\
3  & $ 42.6$ / $21$ ($2.03$) & $ 32.7$ / $22$ ($1.49$) & $ 18.2$ / $19$ ($0.96$) & $ 9.9$ / $22$ ($0.45$) & $ 13.9$ / $20$ ($0.69$)   & $ 22.5$ / $21$ ($1.07$)   \\
4  & $ 16.0$ / $18$ ($0.89$) & $ 19.3$ / $21$ ($0.92$) & $ 21.3$ / $19$ ($1.12$) & $ 24.4$ / $22$ ($1.11$) & $ 12.3$ / $19$ ($0.65$)  & $ 16.7$ / $21$ ($0.79$)   \\
5  & $ 20.0$ / $22$ ($0.91$) & $ 22.4$ / $21$ ($1.06$) & $ 43.9$ / $20$ ($2.19$) & $ 15.4$ / $21$ ($0.73$) & $ 20.7$ / $17$ ($1.22$)  & $ 17.8$ / $20$ ($0.89$)   \\
6  & $ 17.4$ / $22$ ($0.79$) & $ 31.1$ / $21$ ($1.48$) & $ 19.6$ / $18$ ($1.09$) & $ 22.9$ / $21$ ($1.09$) & $ 25.5$ / $19$ ($1.34$)  & $ 15.3$ / $20$ ($0.77$)   \\
7  & $ 21.6$ / $20$ ($1.08$) & $ 34.3$ / $21$ ($1.63$) & $ 24.4$ / $18$ ($1.36$) & $ 27.8$ / $19$ ($1.46$) & $ 19.0$ / $18$ ($1.06$)  & $ 22.6$ / $19$ ($1.19$)   \\
8  & $ 16.7$ / $16$ ($1.04$) & $ 23.2$ / $20$ ($1.16$) & $ 14.6$ / $18$ ($0.81$) & $ 25.2$ / $20$ ($1.26$) & $ 22.1$ / $14$ ($1.58$)  & $ 18.9$ / $18$ ($1.05$)   \\
9  & $ 33.5$ / $21$ ($1.60$) & $ 18.7$ / $21$ ($0.89$) & $ 23.5$ / $17$ ($1.38$) & $ 18.0$ / $20$ ($0.90$) & $ 28.1$ / $17$ ($1.66$)    & $ 24.8$ / $19$ ($1.30$)    \\
10  & $ 23.6$ / $20$ ($1.18$) & $ 22.7$ / $20$ ($1.14$) & $ 23.0$ / $16$ ($1.44$) & $ 19.8$ / $20$ ($0.99$) & $ 10.0$ / $17$ ($0.59$) & $ 31.7$ / $18$ ($1.76$)   \\
11  & $ 21.5$ / $19$ ($1.13$) & $ 21.3$ / $18$ ($1.18$) & $ 30.0$ / $20$ ($1.50$) & $ 29.1$ / $19$ ($1.53$) & $ 17.8$ / $19$ ($0.94$)  & $ 20.0$ / $21$ ($0.95$)   \\
12  & $ 12.0$ / $14$ ($0.86$) & $ 34.1$ / $20$ ($1.71$) & $ 25.3$ / $19$ ($1.33$) & $ 20.5$ / $21$ ($0.98$) & $ 19.0$ / $18$ ($1.06$) & $ 12.1$ / $19$ ($0.64$)   \\
13  & $ 23.2$ / $21$ ($1.11$) & $ 34.3$ / $22$ ($1.56$) & $ 9.7$ / $18$ ($0.54$) & $ 30.4$ / $20$ ($1.52$) & $ 27.3$ / $18$ ($1.52$)  & $ 29.4$ / $20$ ($1.47$)   \\
14  & $ 37.3$ / $20$ ($1.86$) & $ 22.4$ / $21$ ($1.07$) & $ 21.7$ / $18$ ($1.20$) & $ 22.4$ / $19$ ($1.18$) & $ 21.7$ / $17$ ($1.28$)  & $ 24.2$ / $18$ ($1.34$)   \\
15  & $ 30.8$ / $20$ ($1.54$) & $ 14.4$ / $20$ ($0.72$) & $ 19.0$ / $16$ ($1.18$) & $ 18.6$ / $18$ ($1.03$) & $ 24.7$ / $16$ ($1.55$) & $ 26.3$ / $18$ ($1.46$)   \\
16  & $ 26.4$ / $20$ ($1.32$) & $ 25.8$ / $19$ ($1.36$) & $ 14.3$ / $16$ ($0.89$) & $ 20.3$ / $18$ ($1.13$) & $ 33.5$ / $16$ ($2.10$)  & $ 16.0$ / $16$ ($1.00$)    \\
17  & $ 17.8$ / $18$ ($0.99$) & $ 31.1$ / $19$ ($1.64$) & $ 14.1$ / $14$ ($1.01$) & $ 8.7$ / $18$ ($0.48$) & $ 30.3$ / $16$ ($1.90$)   & $ 10.7$ / $16$ ($0.67$)   \\
18  & $ 36.0$ / $18$ ($2.00$) & $ 27.1$ / $19$ ($1.43$) & $ 8.1$ / $14$ ($0.58$) & $ 18.7$ / $18$ ($1.04$) & $ 8.2$ / $15$ ($0.55$)    & $ 7.2$ / $16$ ($0.45$)    \\
19  & $ 36.6$ / $18$ ($2.04$) & $ 17.7$ / $18$ ($0.98$) & $ 21.7$ / $15$ ($1.45$) & $ 12.3$ / $16$ ($0.77$) & $ 16.8$ / $16$ ($1.05$) & $ 17.9$ / $15$ ($1.19$)   \\
20  & $ 24.6$ / $17$ ($1.45$) & $ 14.2$ / $18$ ($0.79$) & $ 32.9$ / $15$ ($2.19$) & $ 14.0$ / $16$ ($0.88$) & $ 18.2$ / $14$ ($1.30$)  & $ 10.1$ / $14$ ($0.72$)   \\
21  & $ 13.8$ / $16$ ($0.86$) & $ 26.6$ / $18$ ($1.48$) & $ 17.0$ / $12$ ($1.42$) & $ 26.0$ / $16$ ($1.62$) & $ 10.0$ / $14$ ($0.71$) & \\
22  & $ 21.1$ / $16$ ($1.32$) & $ 16.3$ / $17$ ($0.96$) & $ 18.5$ / $14$ ($1.32$) & $ 18.5$ / $16$ ($1.16$) & $ 20.2$ / $14$ ($1.44$) & \\
23  &   & $ 17.2$ / $18$ ($0.95$) &   &   & $ 16.9$ / $11$ ($1.54$) & \\
24  &   & $ 18.9$ / $18$ ($1.05$) &   &   &  &  \\
25  &   & $ 24.1$ / $17$ ($1.42$) &   &   &  &  \\
26  &   & $ 27.2$ / $17$ ($1.60$) &   &   &   & \\
27  &   & $ 16.8$ / $16$ ($1.05$) &   &   &  & \\
  \hline
  \noalign{\vskip 1ex}
Total   & $ 541.3$ / $420$ ($1.29$) & $ 657.4$ / $524$ ($1.25$) &$ 453.7$ / $375$ ($1.21$) & $ 440.0$ / $423$ ($1.04$) &$ 448.6$ / $384$ ($1.17$) & $392.3$ / $371$ ($1.06$) \\
  \hline
\end{tabular}
\begin{center}
      Notes: 
      Each column reports the total $\chi^2$ / d.o.f., and the value in the parentheses is the reduced $\chi^2_{\mathrm{r}}$.
       For the 4U 1702$-$429 the best-fit values are $M=1.57~\Msun$, $R=12.17~\mathrm{km}$, $D=5.47~\mathrm{kpc}$ and $X=0.10$.
    In the case of synthetic data we have $M=1.5~\Msun$, $R=12~\mathrm{km}$, $D=6~\mathrm{kpc}$ and $X=0$.
   When computing the $\chi^2$ values we require that the number of counts in each spectral energy bin exceeds $50$.
\end{center}
\end{center}
\end{small}
\end{table*}

\begin{figure*}
\centering
\includegraphics[width=9cm]{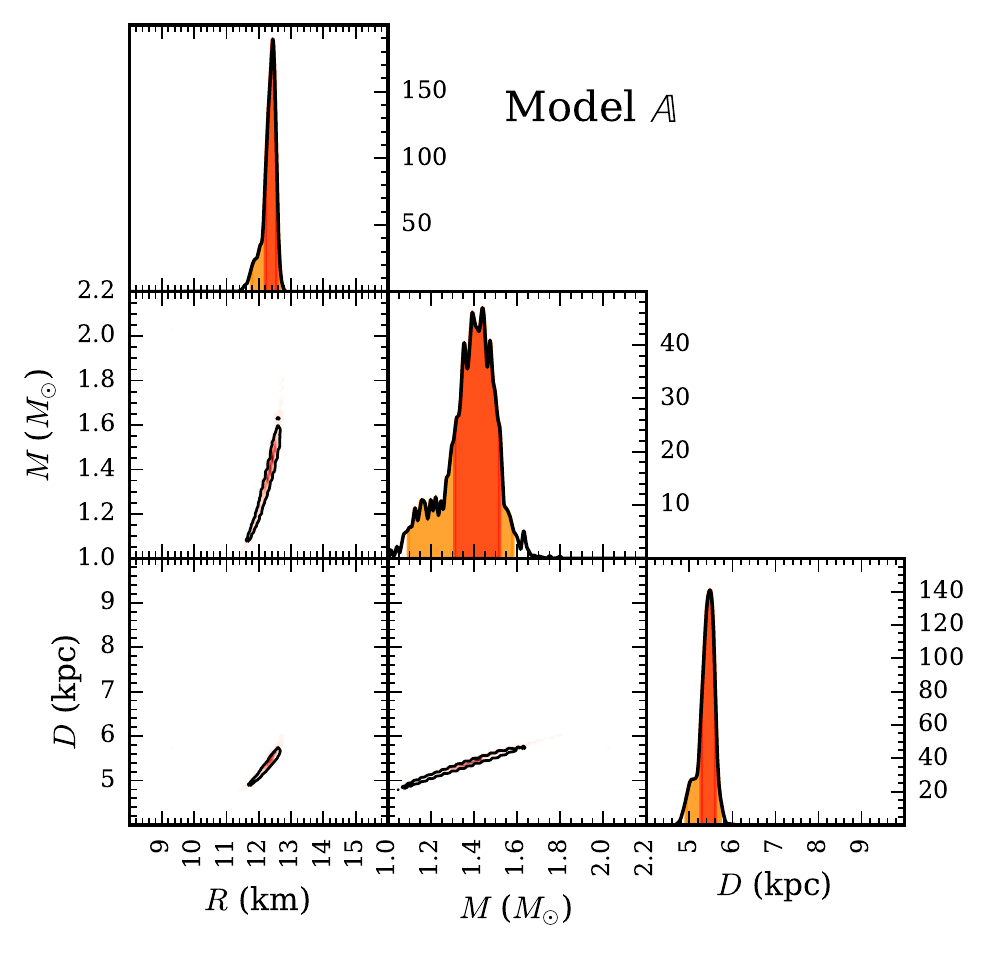}
\includegraphics[width=9cm]{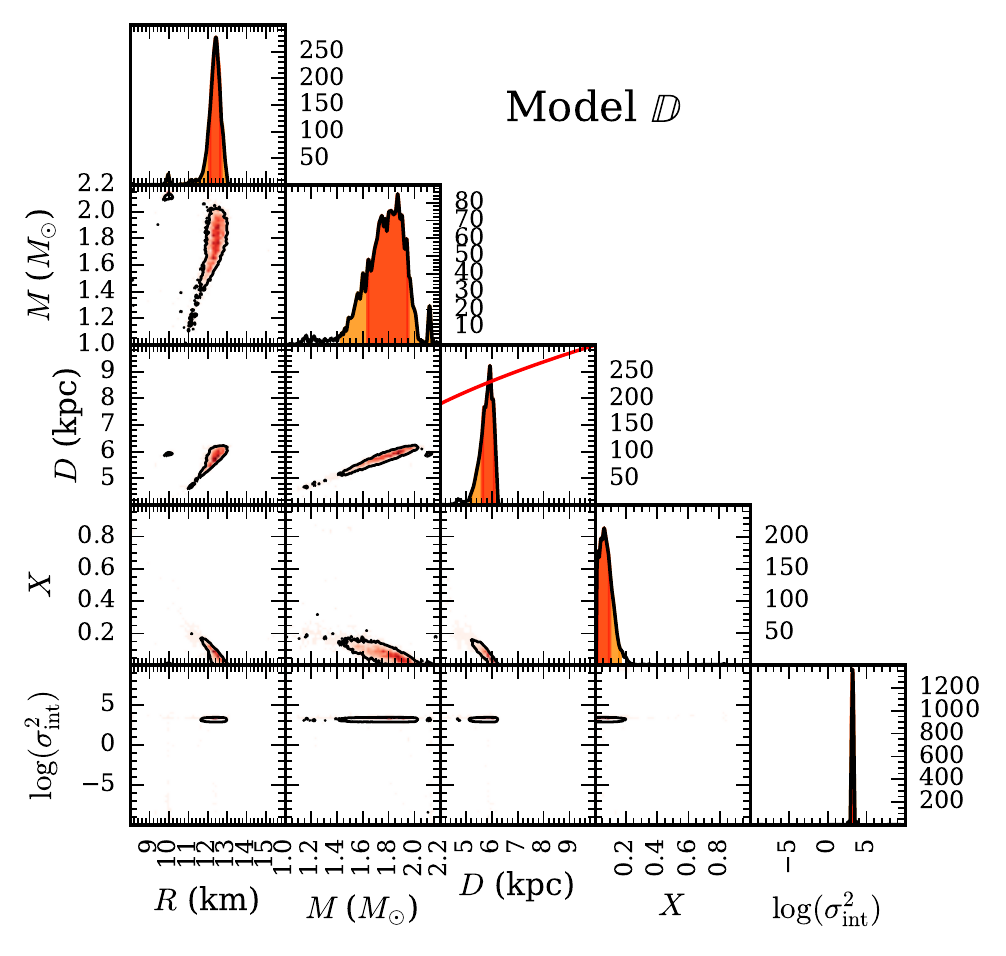}
\caption{\label{fig:1702_tri}
Posterior distributions for the MCMC runs with real data for five PRE bursts from 4U 1702$-$429.
The panels and symbols are the same as in Figs.~\ref{fig:synt_tri} and  \ref{fig:synt07_tri}. 
The red solid line in the $D$ panel shows the prior distribution ($\sqrt{D}$) that we used.
Both models are seen to produce posterior shapes that are similar to what we found in our synthetic data fits (Fig.~\ref{fig:synt07_tri}).
Both models also produce consistent estimates for the radius and distance.
}
\end{figure*}

\begin{figure*}
\centering
\includegraphics[width=8cm]{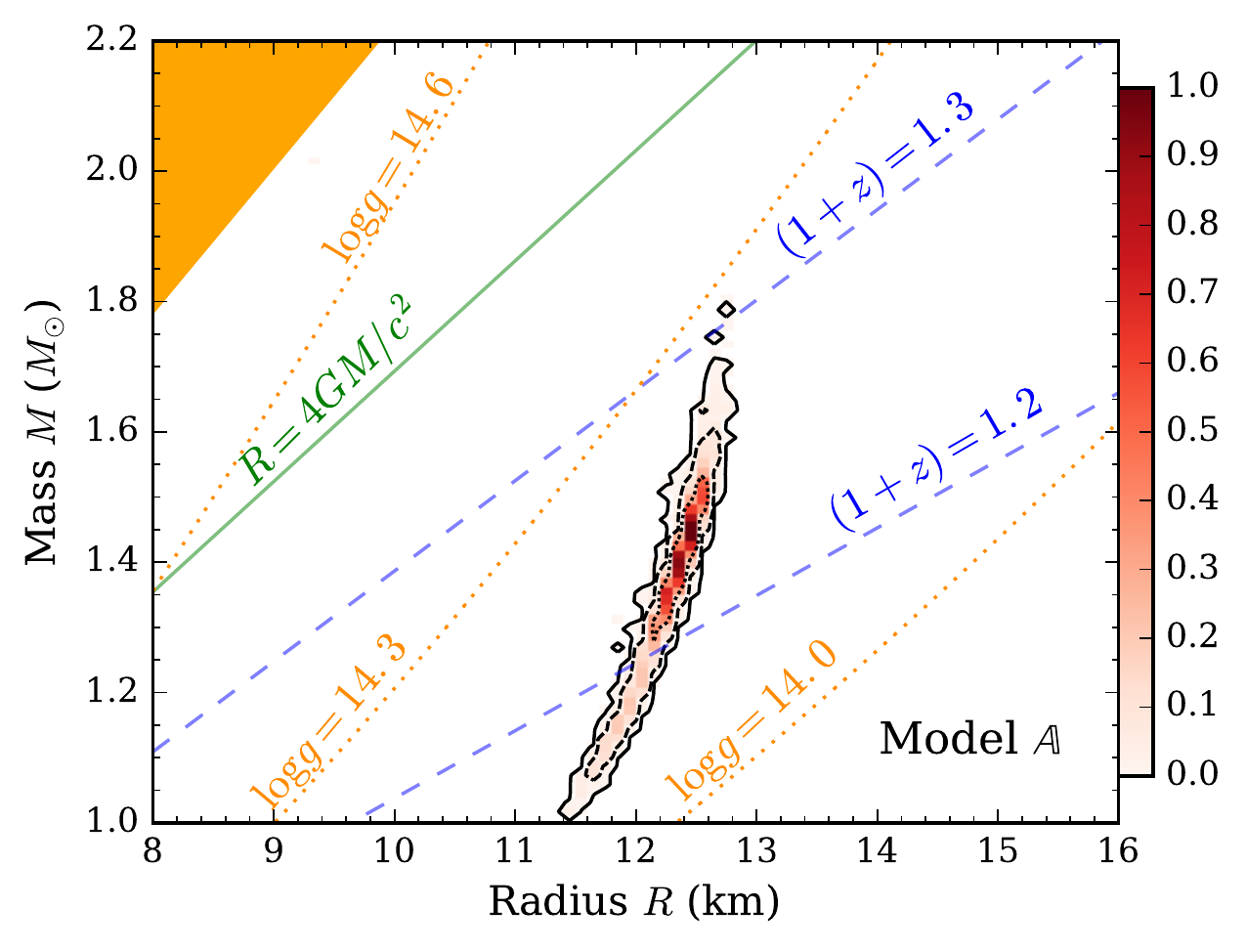}
\includegraphics[width=8cm]{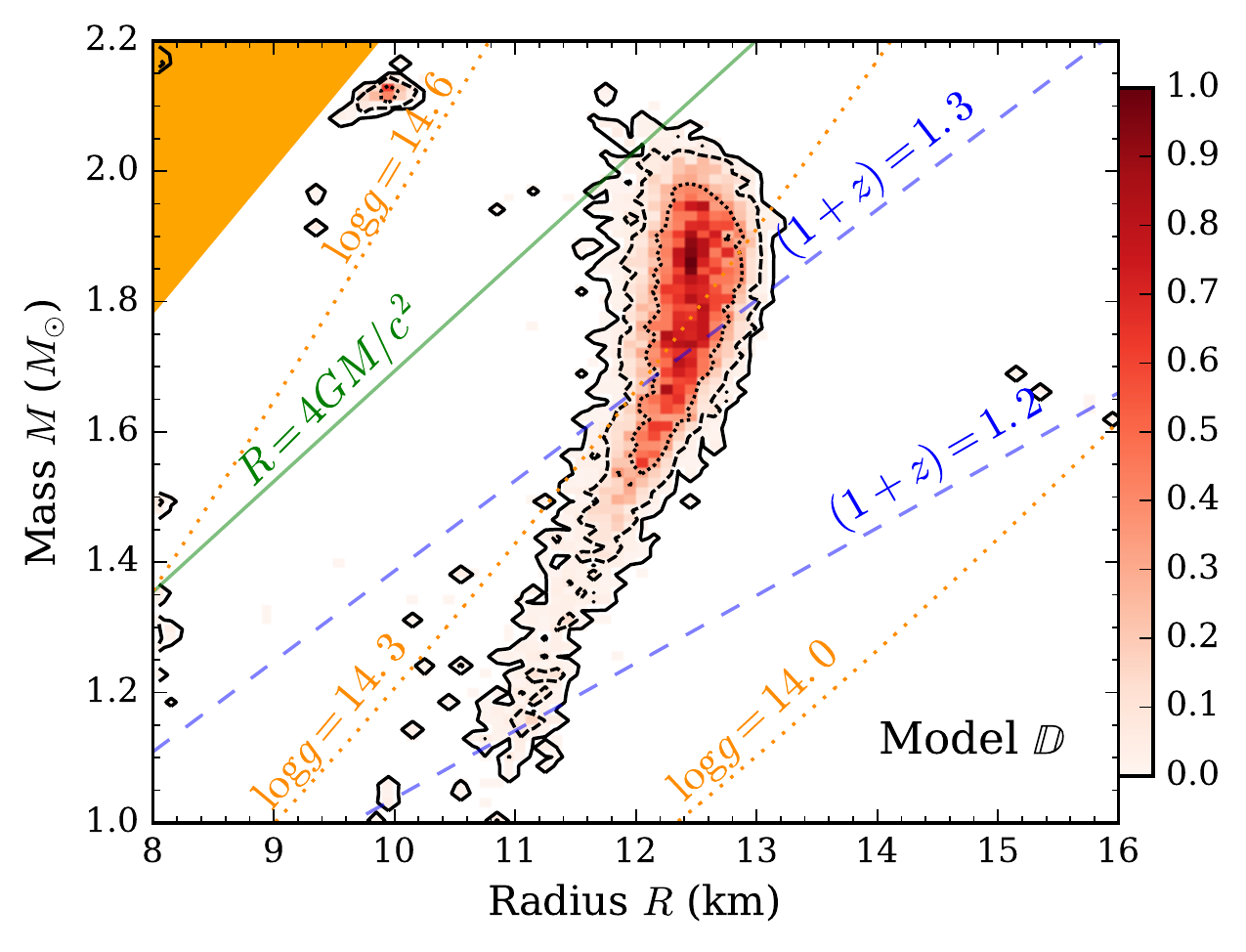}
\caption{\label{fig:1702}
Mass and radius posteriors for 4U 1702$-$429.
The left panel shows the results for Model \modelf{A}, which has fixed $S_{\mathrm{f}}=1$ and $X=0$.
The right panel shows the results for Model \modelf{D}, which has free $X$ and fixed $S_{\mathrm{f}}=1$.
The symbols and legends are the same as in Fig.~\ref{fig:synt_mr}. 
    We recall that for our synthetic data fits (shown in Fig.~\ref{fig:synt07_mr}), Model \modelf{A} underestimates the mass slightly. 
    For $X \ne 0$ the correct mass is obtained using Model \modelf{D}. 
    This suggests that the true mass of 4U 1702$-$429 is in the range $M = 1.4 - 2.0~\Msun$.
    Both Models \modelf{A} and \modelf{D} are capable of recovering the radius used to construct the synthetic data, which suggests that the radius of 4U 1702$-$429 is $R=12.0-12.9~\mathrm{km}$ at $68\%$ credibility.
}
\end{figure*}

As a final test of our atmosphere model goodness-of-fit we can study possible systematic deviations from the model spectra in an individual energy channel level.  In Fig.~\ref{fig:ratio} we show the ratio of the data and the best-fit model flux (from the Model \modelf{D} run) together with a channel-specific $\Delta\chi$ value (defined as $(\mathcal{D} - \mathcal{M})/\sigma$, i.e., difference between data $\mathcal{D}$ and model $\mathcal{M}$ in units of standard error $\sigma$) for each of the five analyzed bursts. 
Additionally, Table~\ref{tab:chis50} shows the sum of $\chi^2$ values for the best-fit models. 
There do not appear to be any significant persistent structures in the residuals. 
There is a slight deficit of flux around $10-14~\mathrm{keV}$ in the burst tails, which corresponds to a $\sim 5\%$ difference between the model and the data ($\Delta \chi \sim 0.4$).  No such features are seen in the synthetic data fits. 
It is therefore possible that this deficit has a physical origin.
We discuss these deviations further in Sect.~\ref{sect:disc}. 
Our total $\chi^2$ (for Model \modelf{D}) is $3103.5$ with $2438$ degrees of freedom (reduced $\chi^2_{\mathrm{r}}$ of $1.27$) for the set of all 5 bursts, when all spectral energy bins with counts less than $20$ are ignored.
The quality of the fit is, however, decreased drastically by the low-count channels and if we, instead, select $50$ as our channel count cutoff we obtain $\chi^2 = 2541.0$ with $2122$ degrees of freedom ($\chi^2_{\mathrm{r}} = 1.20$).


For our actual parameter constraints we consider only Models \modelf{A} and \modelf{D}, because Model \modelf{B} produced somewhat biased constraints with synthetic data and Model \modelf{C} has too much freedom in its parameters (particularly with the inclusion of the surface emitting fraction $S_f$ as a free parameter). 
Fig.~\ref{fig:1702_tri} shows the full posterior distributions, and  Fig.~\ref{fig:1702} shows the two-dimensional $M-R$ posterior distributions in more detail, for each model.
The precision of constraints is similar to what it was for the synthetic data.
We find that $R=12.4 \pm 0.4 ~(0.6)~\mathrm{km}$ and $M=1.4 \pm 0.2 ~(0.4)~\Msun$ for model \modelf{A} (which has fixed chemical composition $X=0$), where we list the $68\%$ (and $95\%$ in parentheses) error regions. 
In Model \modelf{D} the constraints are similar: $R = 12.4^{+ 0.3  ~( 0.6)}_{-0.4 ~( 2.6)}~\mathrm{km}$ and $M= 1.9 ^{+ 0.1 ~( 0.3)}_{-0.3  ~( 0.5)}~\Msun$.
We also find that $X < 0.09$ ($0.16$) at $68\%$ ($95\%$) credibility.
The most probable value is $X=0.06$ rather than $X=0$.
We find a distance of $D = 5.5 \pm 0.4 ~(0.7) ~\mathrm{kpc}$ with model \modelf{A} or $D= 5.9 ^{+ 0.2 ~( 0.3)}_{-0.3  ~( 0.8)}~\mathrm{kpc}$ with model \modelf{D}.
Note that in the model \modelf{D} fits, the second, non-physical high-mass, small-radius family of solutions is now mostly located inside the causality region so it is naturally ruled out by physical considerations.
Some small group of solutions, however, remain at $M=2.2~\Msun$ and $R=10~\mathrm{km}$ that then shifts the lower limit of the $95\%$ radius credibility interval down to $10~\mathrm{km}$ instead of $11.4~\mathrm{km}$ that would be obtained by omitting it entirely.
In contrast to the synthetic data, the real bursts also have a non-zero intrinsic scatter of $\lognat \intscat^2 = 3.2$. 
This corresponds to about $30$ counts per second per energy channel. 
In reality, the intrinsic scatter accommodates more local deviations such as the ones between $10-14~\mathrm{keV}$, as seen in Fig.~\ref{fig:ratio}. 
Such an error in the observed counts reflects a $~\sim 1-5\%$ deviation from the model flux, depending on the $\grg$ value (higher $\grg$ corresponds to higher temperature, i.e., larger count rate for which the deviation is closer to the $1\%$ level, whereas the opposite is true for a small $\grg$).

\section{Discussion}\label{sect:disc}

The measurements we present here result from the use of full atmospheric spectral models of thermonuclear X-ray burst cooling, rather than the usual use of diluted blackbody fits.  
This gives us access to additional information, via the surface redshift and the surface gravity.  
Our new method also allows us to validate many of the assumptions that underlie previous work.
For example, we find that the Eddington limit is reached (and exceeded) near the beginning of the cooling tail (see Fig.~\ref{fig:1702_burst}, that shows the fit results for the first burst in our sample; the remaining 4 bursts are almost identical). 
Hence, this is the most direct validation yet (assuming that the atmosphere models are correct) that at least the bursts that we analyze here are photospheric radius expansion bursts.
Moreover, the dependence of the spectral shape evolution on the atmosphere composition allows us, for the first time, to set reliable limits on the hydrogen mass fraction in the photosphere.
This is made possible by the fact that the temperature evolution of the atmosphere is dependent on the composition.

\subsection{Uncertainties and systematic errors}

By fitting the atmosphere models directly to the data, we can also assess the degree to which the models represent the data. 
Although we reiterate our caveats about the use of $\chi^2$, particularly for model comparisons (see also \citealt{chi}), we note that the Model \modelf{D} fit to  4U 1702$-$429 has $\chi^2/{\rm d.o.f.}=2541.0/2122$ (see also Table~\ref{tab:chis50}), whereas for a simple blackbody fit $\chi^2/{\rm d.o.f.}=2716.3/2010$. 
The blackbody fits are obtained using \textsc{xspec} version 12.9.1 \citep{Arn96} with \textsc{bbodyrad} model.
Thus the model atmosphere fit has an additional 112 degrees of freedom, but its $\chi^2$ is $175.5$ smaller.  
This kind of comparison is, however, not strictly fair because individual blackbody best-fits minimize the quoted $\chi^2$ value for channels with more than 50 counts, whereas the results from the hierarchical modeling are obtained from the MCMC chain that deals with full Poisson or Gaussian likelihoods.
The best-fit values are also very susceptible to the exact energy range used for the fit and so the $\chi^2$ values reported here are only indicative.
When an extra $0.5\%$ calibration error is introduced, as is advised for \textit{RXTE} spectral analysis \citep{Shapo12}, the $\chi^2$ value also decreases, in both cases, by about $30$.
The main difference here that we want to emphasize is that blackbody fits involve two parameters per spectrum (temperature and normalization) whereas direct spectral fitting only has $\grg$ as a parameter for each spectrum (this is true for Models \modelf{A}, \modelf{B} and \modelf{D}; for model \modelf{C} the surface emitting fraction is also a free parameter for each spectrum). 
In addition to the individual spectrum parameters, there are 3 to 5 global parameters, and thus the total number of model parameters is $1\times 116 + [3,4]$ (mass, radius, distance, and hydrogen mass fraction in some cases).
This means that, when compared with the blackbody model, the atmosphere model is able to reduce the number of parameters needed from $\sim 230$ to roughly $120$ while retaining comparable accuracy.

Nonetheless, the formal statistical fit is not good, and this is reflected in the nonzero value of \intscat\ in our Model \modelf{D} fits. 
Hence it appears that there are unmodeled effects in the data.  
An obvious candidate for such complications is the rotation of the star.  
However, because 4U 1702$-$429 has a relatively small rotational frequency of 329~Hz \citep{MSS99}, the effects are unlikely to be large for this star. 
Our preliminary studies show that the effect of rotational broadening of the spectrum is strongest at low and high energies, and hence broadening might account for some of the deviations seen at $E \lesssim 3~\mathrm{keV}$ and $E \ga 12~\mathrm{keV}$ in Fig.~\ref{fig:ratio}. 
The impact on the fit quality (i.e., $\chi^2$), on the other hand, is small because rotational broadening tends to smooth out the spectra without producing any sharp features \citep{NP17}.
Hence, disentangling the effects of rotation from the atmosphere model fits based on the fit quality alone is hard.

As is the case for rotational smearing, we cannot detect if there is non-uniform surface emission in the sense that the temperature varies across the surface.
This is because our model \modelf{C}, which has a free surface emitting fraction $S_{\mathrm{f}}$, can only capture effects where some part of the star is partially covered.
There have been detections of burst oscillations from 4U 1702$-$429 that imply a non-uniform surface temperature \citep{MSS99, GMH08,OWG17}.
However, these have been detected only during the soft state bursts from this particular source \citep{OWG17} and so it could be that the effect, if it exists, is small because it cannot be detected using \textit{RXTE} data.

Another possible source of error is the treatment of heavy elements in the atmosphere. 
\citet{NSK15} showed that heavy elements can have a significant impact on burst spectra.  Most importantly, heavy elements produce photoionization edges around $E \sim 9-14~\mathrm{keV}$.  
The metals are likely to originate from nuclear burning during the burst and might be brought to the surface layers of the star by convection \citep{WBS06,MNA11,MZN14}.
The spectral deviations are expected to appear mainly when the metals start to recombine at lower temperatures after the photosphere has cooled down.  
The detection of such features in other sources with longer, more energetic bursts \citep{iZW10, KNP17} implies that they might also play some role in the shorter bursts analyzed here.

A third possible source of deviations involves the persistent (non-burst) emission, which we currently assume to be constant during the fit. 
Some recent studies indicate that this might not be the case even in the hard state \citep{JZC15, DKC16, Kajava17}.  
However, the initial level of persistent emission for the 5 bursts from 4U 1702$-$42 is very low and we expect this effect to not have a very significant impact on the observed radiation. 
A crude estimate can be obtained by varying the background emission, not with the full hierarchical model, but with individual blackbody fits.
In this case, the constraints come from the Eddington limit and from the normalization $K_{\mathrm{tail}}=(R_{\mathrm{bb}}[\mathrm{km}]/D_{10})^2$ in the burst tail.
Here $R_{\mathrm{bb}}$ is the black body radius ($\propto R$) and $D_{10} = D/10~\mathrm{kpc}$.
By varying the background emission with factors ranging from $0.5$ to $2$, we obtain constraints where the Eddington flux is still accurately recovered but the apparent normalization in the tail is decreased or increased, respectively.
The value of the normalization in the tail, in this case, is within 2\% of the original value.
This is consistent with the fact that the persistent emission during the hard state is about 1\% of the Eddington flux, whereas in the tail $F \sim 0.5 F_{\mathrm{Edd}}$ and so a $2\%$ scatter in the normalization $K$ is expected as $F = K T^4$.
The measured radius $R$ scales roughly (for the fixed compactness) as $R \propto T_{\rm Edd,\infty}^{-4} \propto K_{\mathrm{tail}}/F_{\mathrm{Edd}}$ (see, e.g., Eq.~(A9) in \citealt{PNK14}) so such a deviation in the normalization results in a $\sim 2\%$ scatter in the radius.  
This means that uncertainty related to the persistent emission would then translate to about $250~\mathrm{m}$ absolute error in our measured radius.

The final possible source of error is the neutral hydrogen column density.
It only affects the low-energy channels of \textit{RXTE} which can have an impact on the parameters deduced late in the burst tail.
Near the Eddington limit the radiation peaks at $E \sim 10~\mathrm{keV}$ but when the NS cools down, the bulk of the thermal radiation moves to lower energies.
Hence, the effect is similar to the aforementioned persistent emission where the main effect is on the normalization in the tail of the burst.
When the value of $N_{\mathrm{H}}$ is decreased, the modeled low-energy radiation is affected less by the absorption and so the inferred value of the normalization in the tail also decreases because the model flux is now higher.
Unfortunately, it is also in the Rayleigh-Jeans tail that the surface gravity of the atmosphere models has the greatest effect on the emergent spectra.
Similar considerations as in the persistent emission case show that varying $N_{\mathrm{H}}$ by a factor of $2$ leads to an error in $K_{\mathrm{tail}}$ of $5\%$ that then translates to similar relative error in radius.  
We do, however, note that the measured $N_{\mathrm{H}}$ is usually obtained by other instruments that operate at lower X-ray energies where it is easier to measure the neutral hydrogen column density.
Hence, an uncertainty of a factor of $2$ is certainly overestimating the error related to the value.

Our consideration of error sources leads us to propose that the emission above $F \gtrsim 0.5 F_{\mathrm{Edd}}$ should be the cleanest option for $M-R$ measurements.
However, the high flux near $F \sim F_{\mathrm{Edd}}$ is not free of problems:
early in the cooling tail the count rates from the source are highest and so the detector is affected by the deadtime correction the most.
Deadtime correction near the peak can be as high as $5\%$, which would directly translate to error in the measured $F_{\mathrm{Edd}}$.
This would again translate to similar uncertainty in the radius.
In reality, of course, the deadtime correction scheme proposed by the instrument calibration team should be quite effective at covering this effect and so errors as large as $5\%$ originating from this are not expected.
To be safe, the fluxes between $(0.5 - 0.95) F_{\mathrm{Edd}}$ should give the most stringent constraints.
This, however, decreases the amount of available data even more and for example here in this work we pushed the aforementioned limits to cover $\grg =(0.2 - 0.98)$ that we still think are viable.
Another option would be to try to model the varying background emission and also marginalize over some plausible hydrogen column density range to capture all of the known error sources.
It would be useful to perform such an analysis in the future.
All in all, this shows that we are approaching the absolute measuring accuracy of the \textit{RXTE} satellite.

The best-fit results are also robust against any systematic calibration error in the flux normalization.
Because the constraints for $R$ mainly originate from $F_{\mathrm{Edd}}$ and $K_{\mathrm{tail}}$ (in contrast to the redshift $1+z$ and surface gravity $g$, which have a much weaker effect), the radius is mainly constrained by the temperature evolution of the burst only.
For an unknown systematic energy-independent shift $\zeta$ affecting the observed spectra we still obtain $R \propto \zeta K_{\mathrm{tail}} / F_{\mathrm{Edd}} \propto  \zeta K_{\mathrm{tail}}/ \zeta K T_{\mathrm{Edd},\infty}^4 \propto 1/T_{\mathrm{Edd},\infty}^4$, where $T_{\mathrm{Edd},\infty}$ is the Eddington temperature (see Eq. A9 in \citealt{PNK14} and Eq.~\eqref{eq:Feddobs}). 
This is a distance-independent quantity which makes the derived radius independent of any normalization factor $\zeta$ affecting the observed flux.

\subsection{Comparison and robustness of the constraints}

It is also interesting to compare our analysis of the 4U 1702$-$429 bursts to previous constraints that were obtained using the cooling tail method.  
By applying the cooling tail method to the same set of hard-state bursts that we analyze here, \citet{NSK16} measured the NS radius to be $R\approx 13~\mathrm{km}$ for $M=1.5~\Msun$.  
However, their $M-R$ posteriors have a complicated banana-like shape (see figure 4 in \citealt{NSK16}) and thus the inferred radius depends strongly on the assumed mass.  \citet{NSK16} found that introducing priors on the EoS leads to better constraints on the mass.  
That is, the assumption that all of the sources analyzed in \citet{NSK16} (in addition to 4U 1702$-$429, they used 4U 1724$-$307 and SAX J1810.8$-$2609) originate from the same underlying EoS helps pin down the mass.  
They find that at $68\%$ probability, $M=1.8 \pm 0.3~\Msun$ and $R=11.9 \pm 0.6~\mathrm{km}$, assuming no phase transitions (QMC+Model A in their paper).  
These constraints are in a good agreement with the values derived here, which from Model \modelf{D} are $M \approx 1.9~\Msun$ and $R \approx 12.4~\mathrm{km}$. 
The compositions are also in good agreement: \citet{NSK16} assumed a pure helium composition ($X=0$), whereas here the fit itself shows that $X<0.09$ ($68\%$). 
The distance constraints also agree well: $D = 5.6\pm 0.9~\mathrm{kpc}$ versus $5.5 \pm 0.4~\mathrm{kpc}$, for the cooling tail and the direct spectral fitting methods (model \modelf{A}), respectively.

Note that our $M-R$ results are located away from the critical radius $R=4GM/c^2$ \citep[see][for discussion]{OP15}.  If the constraints from $F_{\mathrm{Edd}}$ and $A$ ($\propto K_{\mathrm{tail}}$) are not consistent with each other,\footnote{Together $F_{\mathrm{Edd}}$ and $A$ set the normalization of the model spectra because both depend on the distance. They both rely on the assumption that it is only the NS surface that is emitting.  If this assumption is invalid, then the observed values might not coincide with the theoretical values.}
then the $M-R$ solution is forced to obey this relation because no real solution exists.  This could happen, for example, if the model is applied to data that it does not describe, such as data from soft-state bursts where the behaviour of the cooling tail might not be totally determined by the NS surface alone \citep[see][]{SLB10, PNK14, KNL14, NSK16}.  

If the true values of $M$ and $R$ are close to the $R=4GM/c^2$ relation, then it is very hard to distinguish this correct solution from an incorrect solution that is forced upon the system by a model that is inconsistent with the data.  Hence, if all of the $M-R$ solutions from multiple sources are located only at this line \citep[see][for such a situation]{OPG16}, it either means that the model is applied inconsistently to data that it does not describe, or that all NSs happen to have the same compactness $M/R = c^2/4G$.

Another interesting aspect of our method is its ability to constrain the composition of the atmosphere.
As can be seen from the synthetic data fits, it is possible to set limits for $X$ with about $10\%$ precision at $68\%$ credibility.
This opens up a whole new window to the study of accretion physics because we can correlate the burst behavior against the composition of the accreted matter.
One should, however, note that the composition we probe here is the composition during the burst, and so in theory the nuclear reprocessing might change the true composition during the measurement.

Although $X=0$ is still consistent with the data, we can ask whether other aspects of the 4U~1702$-$429 bursts are consistent with there being some hydrogen in the atmosphere.  One such consistency check involves the ratio of the fluence of the persistent emission between bursts to the burst fluence itself.  
This ratio, which is usually called $\alpha$, is a measure of the ratio of the gravitational specific energy release to the thermonuclear energy release; because hydrogen fusion releases much more energy than helium fusion, we expect $\alpha$ to be larger when there is less hydrogen present.  
For 4U~1702$-$429, $\alpha \approx 75$ \citep{GMH08}, whereas for 4U 1820$-$303 $\alpha$ is in the range 125--155 \citep{HSW87}. 
The neutron star in 4U~1820$-$303 is usually assumed to have a nearly pure helium atmosphere \citep{Cumming03} so the lower value of $\alpha$ in 4U 1702$-$429 is consistent with the presence of some hydrogen in the latter source.  
Note, however, that the values of $\alpha$ quoted here are the minimum values and might change from burst to burst.
Bursts from 4U 1820$-$303 also exhibit a fast rise and have short timescales, both of which are believed to be consequences of fast helium burning, whereas the 4U 1702$-$429 bursts have longer durations and also longer rise times \citep{GMH08}.  Neither of these findings is conclusive but they do point into the same direction: there should be some traces of hydrogen in the 4U 1702$-$429 atmosphere, as is suggested by our analysis.

\subsection{Future prospects}

It is interesting to consider additional possibilities that are suggested by our new and detailed analysis.  One obvious extension is to include the PRE phase in our fitting.  To do this, however, we will need accurate atmosphere models of extended NS photospheres.  
The advantage would be to increase the available data for analysis, and it should also significantly improve the measurements of $M$ and $R$, because the expansion must be heavily dependent on the redshift $z$ and surface gravity $g$.  
Preliminary work into this direction has already been reported in \citet{MSC16}.  
Their work also allows an independent validation of our atmosphere models. 
Our results agree well with theirs in the range $\grg \approx 0.2 - 1$ implying that, at least in the context of the mutual assumptions set by both computations, the results are reproducible.

Another important, but computationally very expensive, future prospect is to fit all possible X-ray bursts to obtain $M-R$ constraints.  This would help to set groundbreaking constraints on the EoS of the dense matter.  Preliminary studies already validate the previous results that the atmosphere models are not applicable to the soft state bursts \citep{PNK14,KNL14}.  Lastly, it is also important to understand why the models do not agree completely with the data and the physical origin of these deviations.
For this, more work is needed in order to understand the physics and environments of the bursts better.

Despite the uncertainty about the mass, we have improved significantly the constraints on the compactness of the neutron star.
Our mass and radius measurements are encouragingly consistent with recent theoretical analyses of the EoS of cold dense matter \citep{LP16}.
Such a result give hope on the possibility of using astrophysical neutron star measurements to constrain better the behavior of the ultra-dense matter.

\section{Summary}\label{sect:summary}

We have presented the first direct atmosphere model spectral fits to thermonuclear X-ray burst cooling tails.  Our method is a generalization of previous work, in which black body parameters were used as a proxy to trace the  evolution of the energy spectrum.  By fitting the atmosphere models directly to the data we are able to extract more information from the data and also to test some of the physical assumptions made in previous analyses.

We find that fits to synthetic data, which are generated using the same model that we employ for our analyses, reveal as expected a lack of bias and also show the prospects for precise measurements of the mass and radius.  When we apply our fitting procedure to {\it RXTE} data from five hard-state type-I X-ray bursts from 4U~1702$-$429 in Sect.~\ref{sect:1702}, the resulting posteriors bear a clear similarity to the synthetic data, although the formal quality of the fits is worse than in the ideal case.  When we artificially add intrinsic noise in our analysis of the 4U~1702$-$429 data to produce a formally good fit, we find that the radius is constrained to be $R=12.4 \pm 0.4~\mathrm{km}$ at $68\%$ credibility, for both models we employ. 
The source distance is constrained to be between $5.1~\mathrm{kpc} < D < 6.2~\mathrm{kpc}$ ($68\%$ combined credibility limits from model \modelf{A} and \modelf{D}).
We find that the hydrogen mass fraction $X$ for 4U~1702$-$429 can be constrained to $X<0.09$ at 68\% credibility. 
The highest-probability value is $X=0.06$ rather than $X=0$.

The mass seems to be the hardest parameter to constrain. 
When we apply our two models to synthetic data, Model \modelf{A} typically underestimates $M$ by $\sim 0.1~\Msun$, whereas Model \modelf{D} shows even stronger underestimation for atmospheres with no hydrogen in them.
When $X>0$, model \modelf{D} is seen to reproduce the mass when applied to synthetic data.
Our analysis of the {\it RXTE} data for 4U~1702$-$429 yields similar results: Model \modelf{A} gives $M=1.4 \pm 0.2~\Msun$, whereas Model \modelf{D} gives $M=1.9 \pm 0.3~\Msun$. 
If a bias similar to what we find when analyzing synthetic data applies to the 4U~1702$-$429 analysis, then the real mass is expected to lie closer to the model \modelf{D} constraints.
We suggest, therefore, that the $95\%$ credible interval of model \modelf{D} is a trustworthy limit, and in this limit we find that the mass for 4U~1702$-$429 lies in the range $1.4 < M/\Msun < 2.2$.

\begin{acknowledgements}
We appreciate the detailed comments provided by the referee, which helped to improve and clarify the paper.
This research was supported by the University of Turku Graduate School in Physical and Chemical Sciences (JN).
MCM was supported in part by NASA NICER grant NNX16AD90G.
AWS was supported by grant NSF PHY 1554876.
JJEK acknowledges support from the ESA research fellowship programme and the Academy of Finland grants 268740 and 295114.
JN and JJEK acknowledge support from the Faculty of the European Space Astronomy Centre (ESAC). 
VFS was supported by the German Research Foundation (DFG) grant WE 1312/48-1 and by the Russian Government Program of Competitive Growth of Kazan Federal University. 
JP thanks  the Foundations' Professor Pool, the Finnish Cultural Foundation and the National Science Foundation grant PHY-1125915 for support. 
The computer resources of the Finnish IT Center for Science (CSC) and the FGCI project (Finland) are acknowledged.
\end{acknowledgements}

\bibliographystyle{aa}

\end{document}